\documentclass{article}

\usepackage{microtype}
\usepackage{graphicx}
\usepackage{subcaption}
\usepackage{booktabs} 

\usepackage{hyperref}

\usepackage[preprint]{icml2026}

\usepackage{algorithm}
\usepackage{algorithmic}

\usepackage{amsmath}
\usepackage{amssymb}
\usepackage{mathtools}
\usepackage{amsthm}

\usepackage{multirow}
\usepackage{makecell}
\usepackage{enumitem}
\usepackage[most]{tcolorbox}
\usepackage{xcolor}
\definecolor{darkgreen}{RGB}{0,100,0}
\usepackage[capitalize,noabbrev]{cleveref}

\theoremstyle{plain}

\theoremstyle{definition}

\theoremstyle{remark}

\usepackage[textsize=tiny]{todonotes}

\icmltitlerunning{Helix: A Dual-Helix Co-Evolutionary Multi-Agent System}

\begin{document}

\twocolumn[

\icmltitle{Helix: A Dual-Helix Co-Evolutionary Multi-Agent System for \\ Prompt Optimization and Question Reformulation}





\icmlsetsymbol{equal}{*}

\begin{icmlauthorlist}
\icmlauthor{Kewen Zhu}{equal,inst1}
\icmlauthor{Han Tian}{equal,inst2}
\icmlauthor{Liping Yi}{inst1}
\icmlauthor{Zhiming Zhao}{inst1}
\icmlauthor{Xiang Li}{inst1}
\icmlauthor{Qinghua Hu}{inst1}
\end{icmlauthorlist}

\icmlaffiliation{inst1}{College of Intelligence and Computing,
    Tianjin University, Tianjin, China}
\icmlaffiliation{inst2}{College of Computer Science,
    Nankai University, Tianjin, China}

\icmlcorrespondingauthor{Liping Yi}{lipingyi@tju.edu.cn}

\icmlkeywords{Dual-Helix Co-Evolutionary,
Multi-Agent System,
Prompt Optimization,
Question Reformulation}

\vskip 0.3in
]



\printAffiliationsAndNotice{\icmlEqualContribution} 


\begin{abstract}
Automated prompt optimization (APO) aims to improve large language model performance by refining prompt instructions. However, existing methods are largely constrained by fixed prompt templates, limited search spaces, or single-sided optimization that treats user questions as immutable inputs. 
In practice, question formulation and prompt design are inherently interdependent: clearer question structures facilitate focused reasoning and task understanding, while effective prompts reveal better ways to organize and restate queries. Ignoring this coupling fundamentally limits the effectiveness and adaptability of current APO approaches.
In this work, we propose a unified multi-agent system (\textbf{Helix}) that jointly optimizes question reformulation and prompt instructions through a structured three-stage co-evolutionary framework. 
Specifically, Helix integrates (1) \emph{planner-guided decomposition} that breaks optimization into coupled question--prompt objectives, (2) \emph{dual-track co-evolution} where specialized agents iteratively refine and critique each other to produce complementary improvements, and (3) \emph{strategy-driven question generation} that instantiates high-quality reformulations for robust inference.
Extensive experiments on 12 benchmarks against 6 strong baselines demonstrate the effectiveness of Helix, achieving up to $3.95\%$ performance improvements across tasks with favorable optimization efficiency. The code will be made publicly available upon acceptance.

\end{abstract}

\begin{figure}[t]
\centering
\includegraphics[width=0.5\textwidth]{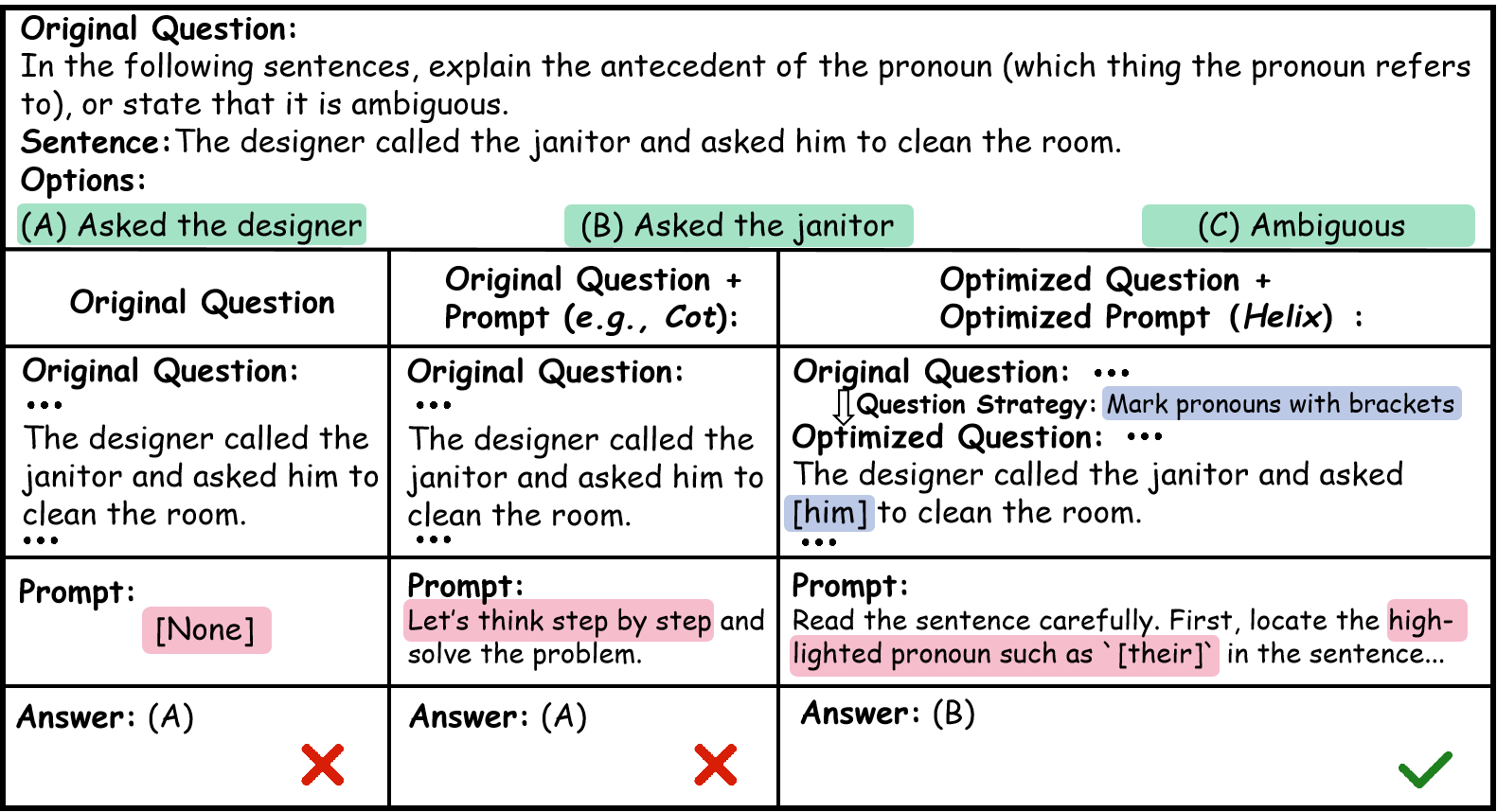}
\caption{Comparison of three strategies for pronoun disambiguation.
Left: original question without prompt instructions yields an incorrect answer.
Middle: adding a CoT prompt to the original question still fails.
Right: Helix jointly optimizes question formulation and prompt instructions, producing the correct prediction.}
\label{fig:figure1}
\end{figure}

\section{Introduction}

Large language models (LLMs) such as GPT-4~\cite{achiam2023gpt} and Llama~\cite{touvron2023llama} have demonstrated strong capabilities across diverse natural language processing tasks. In practice, outputs are jointly determined by the user question and prompt instructions~\cite{brown2020language,wei2021finetuned}, motivating Automated Prompt Optimization (APO)~\cite{pryzant2023automatic} to systematically improve instruction effectiveness beyond manual design.

Recent APO methods mainly follow two paradigms:\textit{ meta-prompting} approaches that employ predefined optimization templates to guide instruction refinement~\cite{yang2023large,ye2024prompt}, and \textit{generate-then-search} methods that construct candidate prompt pools and perform iterative local exploration~\cite{zhou2022large,xu2023reprompting,wang2023promptagent}. Despite their successes, both paradigms exhibit inherent limitations. Meta-prompt approaches rely on fixed templates with limited adaptability across diverse tasks~\cite{yang2023large,ye2024prompt}, performing well in structured domains such as event extraction~\cite{zhang2024semantic} and text-to-symbol transformation~\cite{xu2024symbol} but struggling to accommodate varying reasoning patterns and task-specific optimization strategies. In contrast, generate-search methods restrict exploration to local neighborhoods within predefined prompt clusters~\cite{zhou2022large,xu2023reprompting}, potentially missing globally optimal solutions in the broader prompt space.

Recent advances in LLM-based multi-agent systems have demonstrated strong capabilities for collaborative reasoning and complex problem solving~\cite{tran2025multi,hong2023metagpt,qian2023communicative,ghafarollahi2025sciagents}. Building on these insights, MARS~\cite{zhang2025mars} introduces a planner-guided Teacher--Critic--Student architecture with Socratic dialogue, enabling task-adaptive optimization and broader space exploration for prompts.

However, the entire APO field, including multi-agent approaches, shares a fundamental blind spot: \textbf{existing methods treat questions as direct inputs and optimize only prompt instructions.} This assumption overlooks the inherent interdependence between question formulation and prompt design. Clearer questions enable focused reasoning, while better prompts guide improved question formulation. As shown in Figure~\ref{fig:figure1}, jointly optimizing both dimensions consistently outperforms optimizing either in isolation.


To bridge this gap, we propose a dual-helix co-evolutionary multi-agent system termed \textbf{Helix} that jointly optimizes question formulation and prompt instructions.
(1) A \textit{Planner} first decomposes each task into sequential dual-optimization objectives. 
(2) During training, along two coupled optimization tracks, a \textit{Question-Architect} and a \textit{Prompt-Architect} alternately propose refinements and critique each other’s outputs, forming an interleaved co-evolution process where improvements in one dimension guide updates in the other. A \textit{Mediator} validates whether each refinement achieves complementary progress, producing optimized prompts and transferable question reformulation strategies. 
(3) During inference, the query and question reformulation strategy guide a \textit{Question-Generator} to produce refined questions, while a \textit{Question-Judge} validates their reasonableness, yielding optimized question--prompt pairs for LLM inference.

Our contributions are summarized as threefold:
\begin{itemize}
\item We are the first to formulate automated prompt optimization as a joint question--prompt optimization problem and address it with a dual-helix co-evolutionary multi-agent system.
\item We design a structured three-stage multi-agent optimization framework with planner-guided decomposition, dual-track co-evolution, and strategy-driven question generation with discriminative validation.
\item Extensive experiments on 12 benchmarks against 6 strong baselines demonstrate consistent effectiveness, achieving up to 3.95\% performance improvements across tasks.
\end{itemize}

\section{Related Work}

We review related work in automated prompt optimization and multi-agent-based prompt optimization.

\subsection{Automated Prompt Optimization}

Early prompt optimization explored discrete optimization of hard prompts~\cite{shin2020autoprompt} and continuous optimization of soft prompts~\cite{lester2021power,li2021prefix}. With the emergence of large language models, automated prompt optimization has shifted toward optimizing natural language instructions.

APE~\cite{zhou2022large} initiated this direction by generating and evaluating candidate prompts using LLMs. Subsequent work mainly follows two paradigms. Generate-then-search methods~\cite{zhou2022large,xu2023reprompting,wang2023promptagent,pryzant2023automatic} construct candidate pools and perform local exploration within predefined prompt clusters, which limits global optimization. Meta-prompting approaches~\cite{yang2023large,ye2024prompt} employ predefined optimization templates to guide refinement but lack flexibility across diverse task characteristics.

Overall, existing APO methods focus exclusively on optimizing prompt instructions while treating user questions as fixed inputs, leaving the interaction between question formulation and prompt design largely unexplored.

\subsection{Multi-Agent Prompt Optimization}

LLM-based multi-agent systems decompose complex tasks among specialized agents through collaborative mechanisms~\cite{wu2024autogen,hong2023metagpt,qian2023communicative,tran2025multi}. Key advances include debate mechanisms for mutual critique~\cite{du2023improving,liang2024encouraging}, reflection-based iterative refinement~\cite{shinn2023reflexion}, and meta-agent coordination for high-level planning~\cite{xiong2025mpo}. These paradigms have demonstrated strong performance across domains such as software engineering~\cite{hong2023metagpt,qian2023communicative}, scientific discovery~\cite{ghafarollahi2025sciagents,schmidgall2025agent}, and complex decision-making tasks.

Building on these advances, MARS~\cite{zhang2025mars} applies multi-agent collaboration to prompt optimization through a Teacher--Critic--Student architecture, improving flexibility over template-based methods. However, it maintains fixed agent roles with unidirectional critique and focuses solely on prompt optimization.

In contrast, Helix introduces dual-helix co-evolution via bidirectional critique between \textit{Prompt-Architect} and \textit{Question-Architect}, enabling coordinated optimization of both question formulation and prompt design beyond single-dimension APO.

\begin{figure*}[!t]  
\centering
\includegraphics[width=\textwidth]{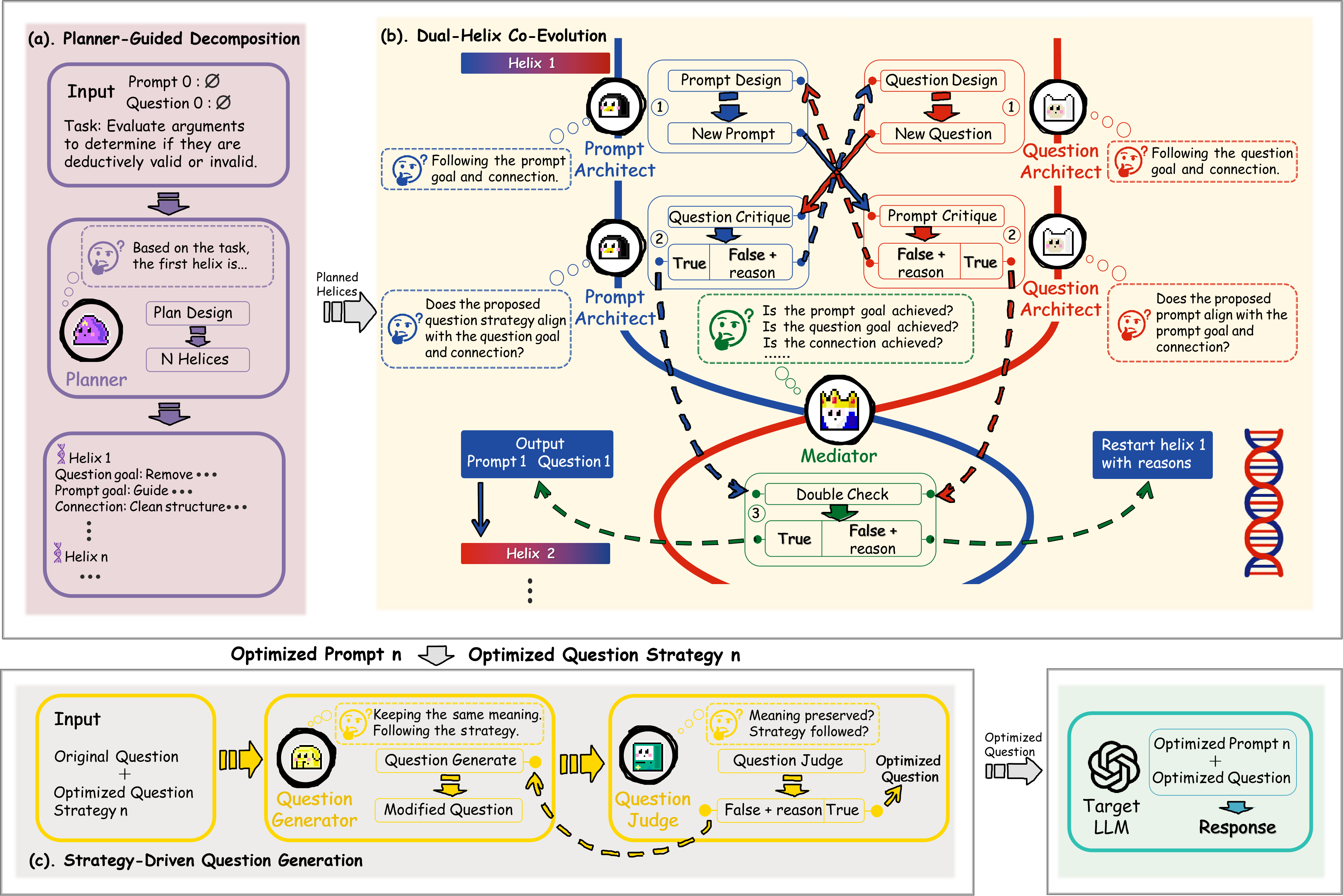}
\caption{Overview of the Helix framework including $6$ LLM-based agents for joint optimization of question reformulation and prompt instructions. 
The \textcircled{1} \textit{Planner} decomposes the task into a sequence of helix objectives, dual-helix co-evolution alternates between \textcircled{2} \textit{Prompt-Architect} and \textcircled{3} \textit{Question-Architect} with \textcircled{4} \textit{Mediator} validation, and the \textcircled{5} \textit{Question-Generator} together with the \textcircled{6} \textit{Question-Judge} produces validated refined questions, which are paired with the optimized prompt and fed to the target LLM for inference.
}
\label{fig:framework}
\end{figure*}

\section{Problem Definition}

Following the definition in prompt optimization~\cite{zhou2022large,yang2023large}, we extend APO to jointly optimize both \textbf{question presentation} and \textbf{prompt instructions}. 
Given a target LLM $\mathcal{M}_{\text{target}}$ and dataset $\mathcal{D}=\{(x,y)\}$, we start from an initial question strategy $Q_0$ and prompt $P_0$, and optimize both to obtain the optimal pair $(Q^*,P^*)$.

We select training examples $\mathcal{D}_{\text{train}}$ for optimization and use $\mathcal{D}_{\text{test}}$ for evaluation. The objective is:

\begin{equation}
(Q^*,P^*)=\arg\max_{Q,P}\sum_{(x,y)\in\mathcal{D}_{{{\text{train}}}}}
f(\mathcal{M}_{\text{target}}(Q(x);P),y),
\end{equation}

where $Q(x)$ denotes the transformed question under strategy $Q$, $P$ is optimized prompt, and $f$ is the performance metric.

Compared with traditional APO, our formulation differs in jointly optimizing $Q$ and $P$ and enforcing their complementarity through bidirectional critique.

\section{The Proposed Helix Framework}

Helix is a dual-helix co-evolutionary multi-agent framework for jointly optimizing question reformulation strategies and prompt instructions. 
The framework operates in two stages: a training stage that learns optimized $(Q^*,P^*)$ through dual-helix co-evolution, and an inference stage that applies them to unseen queries (Figure~\ref{fig:framework}).

\subsection{Planner-Guided Decomposition}

Rather than optimizing questions and prompts in a single step, Helix employs a \textit{Planner} agent $\mathcal{M}_{\text{planner}}$ to decompose the joint optimization task into a sequence of coupled helix objectives. 
Given the task description $\mathcal{T}$ and training examples $\mathcal{D}_{\text{train}}$, the \textit{Planner} generates:

\begin{equation}
\mathcal{H}=[h_1,h_2,\ldots,h_n]=\mathcal{M}_{\text{planner}}(\mathcal{T};\mathcal{D}_{\text{train}}),
\label{eq:planner}
\end{equation}

where each helix $h_i=(g_Q^i,g_P^i,c_i)$ specifies a question reformulation goal $g_Q^i$, a prompt refinement goal $g_P^i$, and a connection $c_i$ describing their intended synergy. 
This decomposition enables progressive and task-adaptive co-evolution across both dimensions.

\subsection{Dual-Helix Co-Evolution}

For each helix objective $h_i\in\mathcal{H}$, Helix initializes the current reformulation strategy and prompt as $(Q_0,P_0)=(Q^{(i-1)},P^{(i-1)})$, where $(Q^{(i-1)},P^{(i-1)})$ denotes the optimized outputs from the previous helix stage (with $Q^{(0)}=P^{(0)}=\emptyset$). The framework then performs iterative debate rounds indexed by $r$.

The \textit{Prompt-Architect} agent $\mathcal{M}_{\text{p-arch}}$ designs an improved prompt draft:

\begin{equation}\label{eq:p-arch-design}
P_r^{\text{draft}}=\mathcal{M}_{\text{p-arch}}^{\text{design}}(h_i;(Q_{r-1},P_{r-1});\phi_{\text{M},r-1};\phi_{\text{Q},r}),
\end{equation}

where $P_{r-1}$ and $Q_{r-1}$ denote the prompt and reformulation strategy from the previous round, $\phi_{\text{M},r-1}$ represents feedback provided by the \textit{Mediator} agent at round $r-1$ (with $\phi_{\text{M},0}=\emptyset$), and $\phi_{\text{Q},r}$ denotes critique feedback from the \textit{Question-Architect} (with $\phi_{\text{Q},r}=\emptyset$ at the start of round $r$).

This draft prompt is then critiqued by the \textit{Question-Architect} $\mathcal{M}_{\text{q-arch}}$:

\begin{equation}\label{eq:q-critique}
(\alpha_{\text{Q},r},\phi_{\text{Q},r})=\mathcal{M}_{\text{q-arch}}^{\text{critique}}(h_i;P_r^{\text{draft}}),
\end{equation}

where $\alpha_{\text{Q},r}\in\{0,1\}$ denotes the acceptance decision (1 for accept and 0 for reject), and $\phi_{\text{Q},r}$ provides critique feedback. If rejected ($\alpha_{\text{Q},r}=0$), the \textit{Prompt-Architect} refines the draft by re-executing Eq.~(\ref{eq:p-arch-design}) with the updated feedback $\phi_{\text{Q},r}$. This critique-refine cycle repeats until acceptance ($\alpha_{\text{Q},r}=1$), yielding the final prompt $P_r$.

Symmetrically, the \textit{Question-Architect} proposes a reformulation strategy draft:

\begin{equation}\label{eq:q-arch-design}
Q_r^{\text{draft}}=\mathcal{M}_{\text{q-arch}}^{\text{design}}(h_i;(Q_{r-1},P_{r-1});\phi_{\text{M},r-1};\phi_{\text{P},r}),
\end{equation}

where $\phi_{\text{P},r}$ denotes critique feedback from the \textit{Prompt-Architect} (with $\phi_{\text{P},r}=\emptyset$ at the start of round $r$). The draft is evaluated by the \textit{Prompt-Architect}:

\begin{equation}\label{eq:p-critique}
(\alpha_{\text{P},r},\phi_{\text{P},r})=\mathcal{M}_{\text{p-arch}}^{\text{critique}}(h_i;Q_r^{\text{draft}}),
\end{equation}

where $\alpha_{\text{P},r}\in\{0,1\}$ denotes acceptance and $\phi_{\text{P},r}$ provides refinement feedback. If rejected ($\alpha_{\text{P},r}=0$), the \textit{Question-Architect} refines the draft by re-executing Eq.~(\ref{eq:q-arch-design}) with the updated feedback $\phi_{\text{P},r}$. This critique-refine cycle repeats until acceptance ($\alpha_{\text{P},r}=1$), yielding the final strategy $Q_r$.

Once both architects accept their respective outputs, the \textit{Mediator} performs joint validation:

\begin{equation}\label{eq:mediator}
([\gamma_{1,r},\gamma_{2,r},\gamma_{3,r}],\phi_{\text{M},r})=
\mathcal{M}_{\text{mediator}}(h_i;(Q_r,P_r)),
\end{equation}

where $\gamma_{\cdot,r}\in\{0,1\}$ indicates pass or fail for three validation dimensions: 
(i) prompt quality satisfying the helix prompt objective $g_P^i$, 
(ii) reformulation quality satisfying the question objective $g_Q^i$, and 
(iii) synergistic consistency fulfilling the connection constraint $c_i$. 
The feedback $\phi_{\text{M},r}$ guides subsequent refinements if any criterion fails.

When all validation dimensions pass, the optimized pair is updated as $(Q^{(i)},P^{(i)})=(Q_r,P_r)$ and the framework advances to the next helix objective. After processing all helix stages, Helix outputs the final optimized pair $(Q^*,P^*)$.

\subsection{Strategy-Driven Question Generation}

During inference, Helix applies the learned reformulation strategy $Q^*$ to unseen queries.

For each test question $x\in\mathcal{D}_{\text{test}}$, the \textit{Question-Generator} $\mathcal{M}_{\text{generator}}$ iteratively produces a refined draft:

\begin{equation}\label{eq:generation}
x_k^{\text{draft}}=\mathcal{M}_{\text{generator}}(x;Q^*;\phi_{\text{J},k-1}),
\end{equation}

where $x_k^{\text{draft}}$ denotes the candidate reformulated question at iteration $k$, and $\phi_{\text{J},k-1}$ is feedback from the \textit{Question-Judge} agent at the previous iteration (with $\phi_{\text{J},0}=\emptyset$).

The draft is then evaluated by the \textit{Question-Judge} $\mathcal{M}_{\text{judge}}$:

\begin{equation}\label{eq:judge}
([\beta_{1,k},\beta_{2,k},\beta_{3,k},\beta_{4,k}],\phi_{\text{J},k})=
\mathcal{M}_{\text{judge}}(x;x_k^{\text{draft}};Q^*),
\end{equation}

where $\beta_{\cdot,k}\in\{0,1\}$ indicates pass or fail for four validation dimensions: 
(i) semantic preservation with respect to the original question $x$, 
(ii) compliance with the learned reformulation strategy $Q^*$, 
(iii) improvement in clarity and structure, and 
(iv) absence of any solving instructions or answer leakage. 
The feedback $\phi_{\text{J},k}$ provides corrective signals when any criterion fails.

When all validation indicators satisfy $\beta_{\cdot,k}=1$, the refined question is accepted as the final optimized input $x^*=x_k^{\text{draft}}$. 
The optimized question is then paired with the evolved prompt $P^*$ and passed to the target LLM:

\begin{equation}\label{eq:prediction}
y_{\text{pred}}=\mathcal{M}_{\text{target}}(x^*;P^*),
\end{equation}

which produces the predicted answer for performance evaluation.
Alg.~\ref{alg:helix} describes the complete process.

Overall, Helix jointly optimizes question reformulation and prompt refinement via planner-guided decomposition, dual-helix co-evolution, and discriminative generation. 
Bidirectional agent debate with \textit{mediator} and judge validation drives complementary improvements and enforces semantic consistency across both dimensions across training and inference stages. 
By systematically exploring the joint question--prompt space, Helix elevates automated prompt optimization from single-dimension tuning to a co-evolutionary multi-agent paradigm.

\begin{algorithm}[t]
\caption{\texttt{Helix}}
\label{alg:helix}
\small
\begin{algorithmic}[1]
\STATE \textbf{Input:} Task $\mathcal{T}$, $\mathcal{D}_{\text{train}}$, $\mathcal{D}_{\text{test}}$, runs $T$.
\STATE \textbf{Output:} Optimized $(Q^*,P^*)$.

\STATE Initialize best score $s^*\gets -\infty$.
\FOR{$t=1$ \textbf{to} $T$}
    \STATE \textbf{Planner-Guided Decomposition:}
    \STATE Generate helix tasks $\mathcal{H}=[h_1,\dots,h_n]$ by Planner (Eq.~\ref{eq:planner}).

    \STATE $(Q^{(0)},P^{(0)})\gets(\emptyset,\emptyset)$

    \FOR{$i=1$ \textbf{to} $n$}
        \STATE \textbf{Dual-Helix Co-Evolution:}
        \REPEAT 
            \STATE \textit{Prompt-Architect} proposes prompt update (Eq.~\ref{eq:p-arch-design}).
            \STATE \textit{Question-Architect} critiques (Eq.~\ref{eq:q-critique}).
            \STATE \textit{Prompt-Architect} refines (Eq.~\ref{eq:p-arch-design}).
            \STATE \textit{Question-Architect} gives question strategy (Eq.~\ref{eq:q-arch-design}).
            \STATE \textit{Prompt-Architect} critiques (Eq.~\ref{eq:p-critique}).
            \STATE \textit{Question-Architect} refines (Eq.~\ref{eq:q-arch-design}).
            \STATE \textit{Mediator} validates joint improvement (Eq.~\ref{eq:mediator}).
        \UNTIL{all validation criteria are satisfied}
        \STATE Update $(Q^{(i)},P^{(i)})$.
    \ENDFOR

    \STATE $(\tilde Q,\tilde P)\gets(Q^{(n)},P^{(n)})$

    \STATE \textbf{Strategy-Driven Question Generation:}
    \FORALL{$x\in\mathcal{D}_{\text{test}}$}
        \REPEAT
            \STATE \textit{Generator} reformulated question using $\tilde Q$ (Eq.~\ref{eq:generation}).
            \STATE \textit{Judge} validates quality and semantics (Eq.~\ref{eq:judge}).
        \UNTIL{validation passes}
        \STATE Predict answer with target model using $\tilde P$ (Eq.~\ref{eq:prediction}).
    \ENDFOR

    \STATE Compute performance score $s_t$.
    \IF{$s_t>s^*$}
        \STATE $(Q^*,P^*)\gets(\tilde Q,\tilde P)$, $s^*\gets s_t$
    \ENDIF
\ENDFOR
\end{algorithmic}
\end{algorithm}

\section{Experiments}

We conduct extensive experiments across 12 diverse benchmarks to evaluate the effectiveness of Helix. 

\subsection{Experimental Setup}

\paragraph{Tasks and Datasets.} 
We evaluate Helix on 12 datasets covering diverse reasoning and knowledge-intensive tasks from established benchmarks, including \textbf{BBH}\footnote{\tiny\url{https://huggingface.co/datasets/SaylorTwift/bbh}}~\cite{suzgun2023challenging}, 
\textbf{MMLU}\footnote{\tiny\url{https://huggingface.co/datasets/cais/mmlu}}~\cite{hendrycks2020measuring}, 
\textbf{MMLU-Pro}\footnote{\tiny\url{https://huggingface.co/datasets/TIGER-Lab/MMLU-Pro}}~\cite{wang2024mmlu}, 
\textbf{LSAT-AR} from AGIEval\footnote{\tiny\url{https://huggingface.co/datasets/hails/agieval-lsat-ar}}~\cite{zhong2024agieval}, 
and \textbf{AQuA-RAT}\footnote{\tiny\url{https://huggingface.co/datasets/deepmind/aqua_rat}}~\cite{ling2017program}.

\paragraph{Baselines.} 
We compare Helix with manual prompting strategies:
\textbf{No Prompt}, 
\textbf{CoT}~\cite{wei2022chain};
APO methods, including 
\textbf{APE}~\cite{zhou2022large}, 
\textbf{OPRO}~\cite{yang2023large}, 
\textbf{PE2}~\cite{ye2024prompt};
and the state-of-the-art multi-agent method \textbf{MARS}~\cite{zhang2025mars}. More details are provided in Appendix~\ref{app:datasets}.

To explicitly analyze the contribution of question reformulation and prompt refinement, we further evaluate four controlled configurations:
\begin{itemize}
    \item \textbf{Q + P-Opt}, where only the prompt is optimized while the original questions are kept unchanged;
    \item \textbf{Q-Opt}, where only the question reformulation strategy is optimized without any prompt;
    \item \textbf{Q-Opt + CoT}, where the optimized questions are paired with a fixed chain-of-thought prompt;
    \item \textbf{Q-Opt + P-Opt}, which corresponds to the full Helix framework jointly optimizing both dimensions of question and prompt through dual-helix co-evolution.

\end{itemize}

\paragraph{Evaluation Metrics.} 
We adopt two evaluation criteria. The primary metric is \textbf{Accuracy} (\%), defined as:

\begin{equation}
\text{Acc}=\frac{|\{(x,y)\in\mathcal{D}_{\text{test}}:y_{\text{pred}}=y\}|}{|\mathcal{D}_{\text{test}}|},
\label{eq:accuracy}
\end{equation}

which measures the proportion of correctly predicted instances.
To evaluate optimization efficiency, we compute \textbf{Prompt Efficiency (PE)}~\cite{zhang2025mars} as:

\begin{equation}
\text{PE}=\frac{\text{Accuracy}}{\text{Consumption}},
\label{eq:pe}
\end{equation}

where Consumption denotes the number of LLM API calls required for prompt optimization. 
For fair comparison, PE is measured under the \textbf{Q + P-Opt} configuration, excluding question generation costs. 
Higher PE indicates better performance with lower optimization overhead.

\paragraph{Implementation Details.} 
All Helix agents are implemented using GPT-4o~\cite{achiam2023gpt}. 
For the target LLM, we use GPT-4o on BBH and Qwen2.5-32B-Instruct~\cite{qwen2025qwen25technicalreport} on the remaining benchmarks. 
We set the number of optimization runs to $T=10$, with at most $R_{\max}=3$ dual-helix co-evolution rounds per helix objective and $K_{\max}=3$ generator--judge iterations during inference. 
We report the best-performing configuration across all runs. 
More details are provided in Appendix~\ref{app:settings}.

\begin{table*}[t]
\centering
\caption{Accuracy (\%) comparison across 12 tasks. 
Helix with dual optimization (\textbf{Q-Opt + P-Opt}) consistently achieves the best performance across all benchmarks. 
Bold denotes the highest result for each task. Abbreviations: D.QA = Disambiguation QA, G.S. = Geometric Shapes, F.F. = Formal Fallacies, R.N. = Ruin Names, S.U. = Sports Understanding, L.A. = LSAT-AR, C.B. = College Biology, E.E. = Electrical Engineering, M.T. = Marketing, Hist = History, Phil = Philosophy, A.R. = AQuA-RAT.}
\setlength{\tabcolsep}{4.8pt}
\renewcommand{\arraystretch}{1.2}
\small
\begin{tabular}{l|ccccc|c|ccc|cc|c|c}
\toprule
\textbf{Models} & \multicolumn{5}{c|}{\textbf{BBH}} & \textbf{AGIEval} & \multicolumn{3}{c|}{\textbf{MMLU}} & \multicolumn{2}{c|}{\textbf{MMLU-Pro}} & \textbf{Math} & \textbf{Avg} \\
\cline{2-6} \cline{7-7} \cline{8-10} \cline{11-12} \cline{13-13}
 & \textbf{D.QA} & \textbf{G.S.} & \textbf{F.F.} & \textbf{R.N.} & \textbf{S.U.} & \textbf{L.A.} & \textbf{C.B.} & \textbf{E.E.} & \textbf{M.T.} & \textbf{Hist} & \textbf{Phil} & \textbf{A.R.} & \\
\hline
\multicolumn{14}{l}{\textit{Manual Prompting}} \\
\hline
No Prompt & 53.20 & 52.40 & 77.40 & 82.80 & 78.80 & 35.09 & 89.58 & 76.55 & 88.32 & 55.12 & 54.11 & 86.61 & 69.17 \\
CoT & 74.80 & 54.80 & 82.00 & 87.20 & 80.00 & 33.22 & 91.36 & 82.07 & 90.06 & 57.74 & 57.31 & 87.40 & 73.16 \\
\hline
\multicolumn{14}{l}{\textit{Automated Prompt Optimization}} \\
\hline
APE & 56.50 & 61.20 & 82.40 & 83.60 & 81.40 & 35.61 & 89.23 & 78.23 & 89.23 & 58.40 & 58.43 & 84.32 & 71.55 \\
OPRO & 50.00 & 46.40 & 84.40 & 87.60 & 78.40 & 37.23 & 91.15 & 81.45 & 90.12 & 58.84 & 59.44 & 85.44 & 70.87 \\
PE2 & 67.20 & 56.00 & 84.00 & 88.40 & 82.40 & 36.46 & 91.24 & 79.44 & 91.54 & 59.25 & 56.43 & 87.32 & 73.31 \\
MARS & 71.20 & 67.60 & 90.20 & 90.80 & 90.20 & 38.09 & 91.36 & 73.79 & 91.88 & 61.32 & 62.63 & 87.88 & 76.41 \\
\hline
\multicolumn{14}{l}{\textit{Helix (Ours)}} \\
\hline
Q + P-Opt & 70.80 & 72.40 & 90.80 & 91.20 & 90.40 & 38.53 & 91.67 & 80.69 & 93.16 & 60.63 & 63.33 & 89.37 & 77.75 \\
Q-Opt & 52.00 & 58.00 & 80.00 & 87.50 & 79.60 & 39.13 & 91.56 & 77.93 & 92.31 & 59.54 & 57.72 & 86.79 & 71.84 \\
Q-Opt + CoT & 70.00 & 60.80 & 86.40 & 90.20 & 82.40 & 38.70 & 92.36 & 80.69 & 92.31 & 60.31 & 58.44 & 90.16 & 75.23 \\
\hline
\textbf{Q-Opt + P-Opt} & \textbf{73.20} & \textbf{73.20} & \textbf{94.05} & \textbf{93.20} & \textbf{92.40} & \textbf{43.48} & \textbf{93.06} & \textbf{85.52} & \textbf{95.23} & \textbf{64.04} & \textbf{65.21} & \textbf{91.73} & \textbf{80.36} \\
\bottomrule
\end{tabular}
\label{tab:main_results}
\end{table*}

\subsection{Overall Performance}

Table~\ref{tab:main_results} summarizes performance across all 12 tasks. 
Helix with full dual optimization (\textbf{Q-Opt + P-Opt}) achieves the highest average accuracy of 80.36\%, consistently outperforming all baselines. 
In particular, it improves over the strongest existing APO method MARS by 3.95\% and manual CoT prompting by 7.20\%.

Optimizing a single dimension already yields clear benefits. 
Prompt-only optimization (\textbf{Q + P-Opt}) reaches 77.75\%, surpassing MARS by 1.34\%. 
Question reformulation alone also brings notable gains: \textbf{Q-Opt} improves direct answering from 69.17\% to 71.84\%, while \textbf{Q-Opt + CoT} improves fixed CoT prompting from 73.16\% to 75.23\%. 
These results indicate that question presentation substantially influences model performance even without prompt refinement.

Nevertheless, jointly optimizing both dimensions consistently delivers the best results. 
Full Helix outperforms the strongest single-dimension configuration by 2.61\% on average, with particularly strong improvements on reasoning-intensive tasks such as BBH and MMLU-related benchmarks. 
This demonstrates the synergistic effect of co-evolving question reformulation and prompt refinement beyond isolated optimization.

\begin{figure}[t]
\centering
\includegraphics[width=0.8\columnwidth]{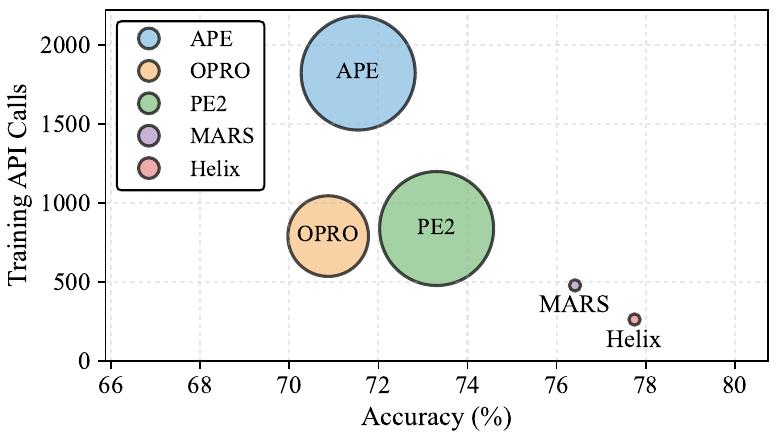}
\caption{Accuracy--cost trade-off across 12 tasks for different prompt optimization methods. Bubble size denotes the number of training samples, with Helix achieving the highest accuracy with the fewest API calls using only a single sample.}

\label{fig:efficiency_bubble}
\vspace{-0.5em}
\end{figure}

\subsection{Optimization Efficiency}

\begin{figure}[t]
\centering
\includegraphics[width=0.9\columnwidth]{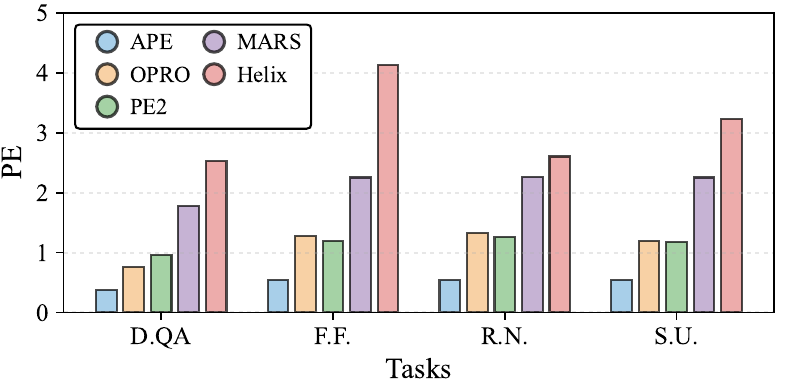}
\caption{Prompt efficiency (PE) comparison on four representative BBH tasks, where Helix consistently achieves the highest performance per optimization cost.}
\label{fig:pe_analysis}
\vspace{-0.4em}
\end{figure}

Figure~\ref{fig:efficiency_bubble} illustrates the trade-off between training cost and accuracy across all 12 tasks. 
Helix achieves higher accuracy while requiring substantially fewer LLM calls than prior APO methods, reducing total optimization cost by approximately 45\% compared to MARS.

This efficiency stems from Helix’s structured co-evolution. 
Dual-role architects reduce agent redundancy, \textit{mediator} validation prunes incompatible updates early, and joint optimization across question and prompt spaces accelerates convergence. 
In practice, most helix objectives converge within a single co-evolution round.

Figure~\ref{fig:pe_analysis} reports Prompt Efficiency (PE) on four BBH tasks. 
Helix consistently achieves the highest PE, with particularly large gains on challenging reasoning tasks such as D.QA. and F.F., while maintaining strong efficiency across R.N. and S.U.

 \begin{table}[t]
 \centering
  \caption{Accuracy (\%) of helix number on MMLU and MMLU-Pro tasks. $\Delta$ denotes the performance drop relative to full Helix.}
 \setlength{\tabcolsep}{2.8pt}
 \renewcommand{\arraystretch}{1.2}
 \small
 \begin{tabular}{l|ccccc}
 \toprule
 \textbf{helix} & \textbf{C.B.} & \textbf{M.T.} & \textbf{E.E.} & \textbf{Hist} & \textbf{Phil} \\
 \hline
 Full & 93.06 & 95.23 & 85.52 & 64.04 & 65.21 \\
 \hline
 1 helix & 85.42 & 87.34 & 77.85 & 55.82 & 56.73 \\
 \quad $\Delta$ & \textcolor{darkgreen}{(-7.64)} & \textcolor{darkgreen}{(-7.89)} & \textcolor{darkgreen}{(-7.67)} & \textcolor{darkgreen}{(-8.22)} & \textcolor{darkgreen}{(-8.48)} \\
 2 helices & 89.58 & 91.45 & 81.38 & 59.11 & 59.88 \\
 \quad $\Delta$ & \textcolor{darkgreen}{(-3.48)} & \textcolor{darkgreen}{(-3.78)} & \textcolor{darkgreen}{(-4.14)} & \textcolor{darkgreen}{(-4.93)} & \textcolor{darkgreen}{(-5.33)} \\
 3 helices & \textbf{93.06} & \textbf{95.23} & 83.79 & 61.47 & 62.05 \\
 \quad $\Delta$ & (0.00) & (0.00) & \textcolor{darkgreen}{(-1.73)} & \textcolor{darkgreen}{(-2.57)} & \textcolor{darkgreen}{(-3.16)} \\
 4 helices & - & - & \textbf{85.52} & \textbf{64.04} & 63.88 \\
 \quad $\Delta$ & - & - & (0.00) & (0.00) & \textcolor{darkgreen}{(-1.33)} \\
 5 helices & - & - & - & - & \textbf{65.21} \\
 \quad $\Delta$ & - & - & - & - & (0.00) \\
 \bottomrule
 \end{tabular}

 \label{tab:helix_ablation}
 \vspace{-0.4em}
 \end{table}

\subsection{Impact of Helix Number $n$}

Table~\ref{tab:helix_ablation} reports performance on MMLU and MMLU-Pro under different numbers of helix objectives. 
We observe clear task-dependent saturation behavior. 
Relatively simpler tasks (C.B., M.T.) reach optimal performance with three helices, moderately complex tasks (E.E., History) benefit from four helices, while the most challenging task (Philosophy) continues improving up to five helices. 

Importantly, the largest performance gains occur in the first one to two helices, with diminishing returns thereafter, indicating efficient early-stage co-evolution. 
These results suggest that the \textit{Planner} agent produces appropriately granular decompositions aligned with task complexity, enabling progressive yet efficient optimization. The saturation pattern also validates that Helix avoids over-optimization, automatically stabilizing when further decomposition yields negligible benefits.

 \begin{table}[t]
 \centering
  \caption{Accuracy (\%) of different training sample sizes $|\mathcal{D}_{\text{train}}|$ on MMLU and MMLU-Pro tasks.}
 \setlength{\tabcolsep}{2.5pt}
 \renewcommand{\arraystretch}{1.15}
 \small
 \begin{tabular}{l|c|ccc|cc|c}
 \toprule
 \textbf{Model} & \textbf{$|\mathcal{D}_{\text{train}}|$} & \textbf{C.B.} & \textbf{E.E.} & \textbf{M.T.} & \textbf{Hist} & \textbf{Phil} & \textbf{Avg} \\
 \midrule
 APE & 100 & 89.23 & 78.23 & 89.23 & 58.40 & 58.43 & 74.70 \\
 OPRO & 50 & 91.15 & 81.45 & 90.12 & 58.84 & 59.44 & 76.20 \\
 PE2 & 100 & 91.24 & 79.44 & 91.54 & 59.25 & 56.43 & 75.58 \\
 MARS & 1 & 91.36 & 73.79 & 91.88 & 61.32 & 62.63 & 76.20 \\
 \midrule
 Helix & 1 & 93.06 & 85.52 & 95.23 & 64.04 & 65.21 & 80.61 \\
 Helix & 2 & 94.44 & 87.24 & 96.58 & 65.47 & 66.76 & 82.10 \\
 Helix & 3 & \textbf{95.14} & \textbf{87.93} & \textbf{97.08} & \textbf{66.33} & \textbf{67.54} & \textbf{82.80} \\
 \bottomrule
\end{tabular}
\vspace{-1.4em}
 \label{tab:sample_size}
 \end{table}

 \subsection{Impact of Training Sample Size $|\mathcal{D}_{\text{train}}|$}

A key advantage of Helix is its low-shot optimization regime, requiring only a few training examples to learn effective question--prompt strategies. 
Table~\ref{tab:sample_size} compares performance under varying training sample sizes.

As $|\mathcal{D}_{\text{train}}|$ increases from 1 to 3, Helix consistently improves across all tasks, indicating that additional examples help refine co-evolutionary optimization. 
However, the performance gains diminish beyond two samples, suggesting rapid convergence with minimal supervision. 

Notably, even the 1-shot setting already surpasses strong baselines such as MARS, demonstrating that Helix can achieve high performance with extremely limited optimization data. 
This highlights the practicality of Helix in resource-constrained scenarios, where collecting large optimization sets is infeasible.

\begin{figure}[t]
\centering
\includegraphics[width=0.5\textwidth]{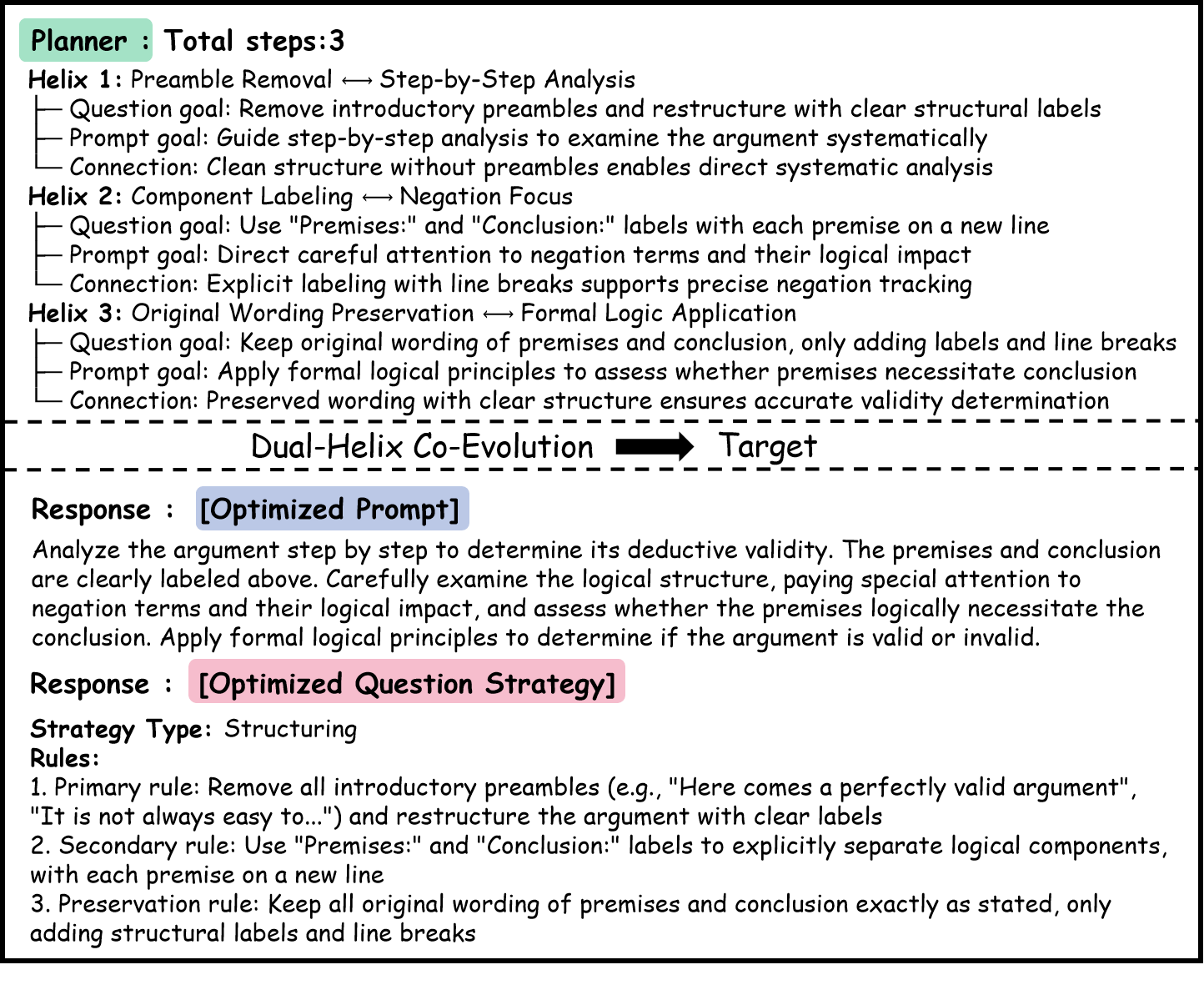}

\caption{Training-stage dual-helix co-evolution on the Formal Fallacies task. 
The \textit{Planner} decomposes the task into sequential helix objectives with coupled question reformulation and prompt refinement goals. 
\textit{Prompt-Architect} and \textit{Question-Architect} iteratively critique and refine each other under \textit{Mediator} validation, yielding aligned optimized prompts and transferable question reformulation strategies.}
\label{fig:case_study}

\end{figure}

\begin{figure}[t]
\centering
\includegraphics[width=0.5\textwidth]{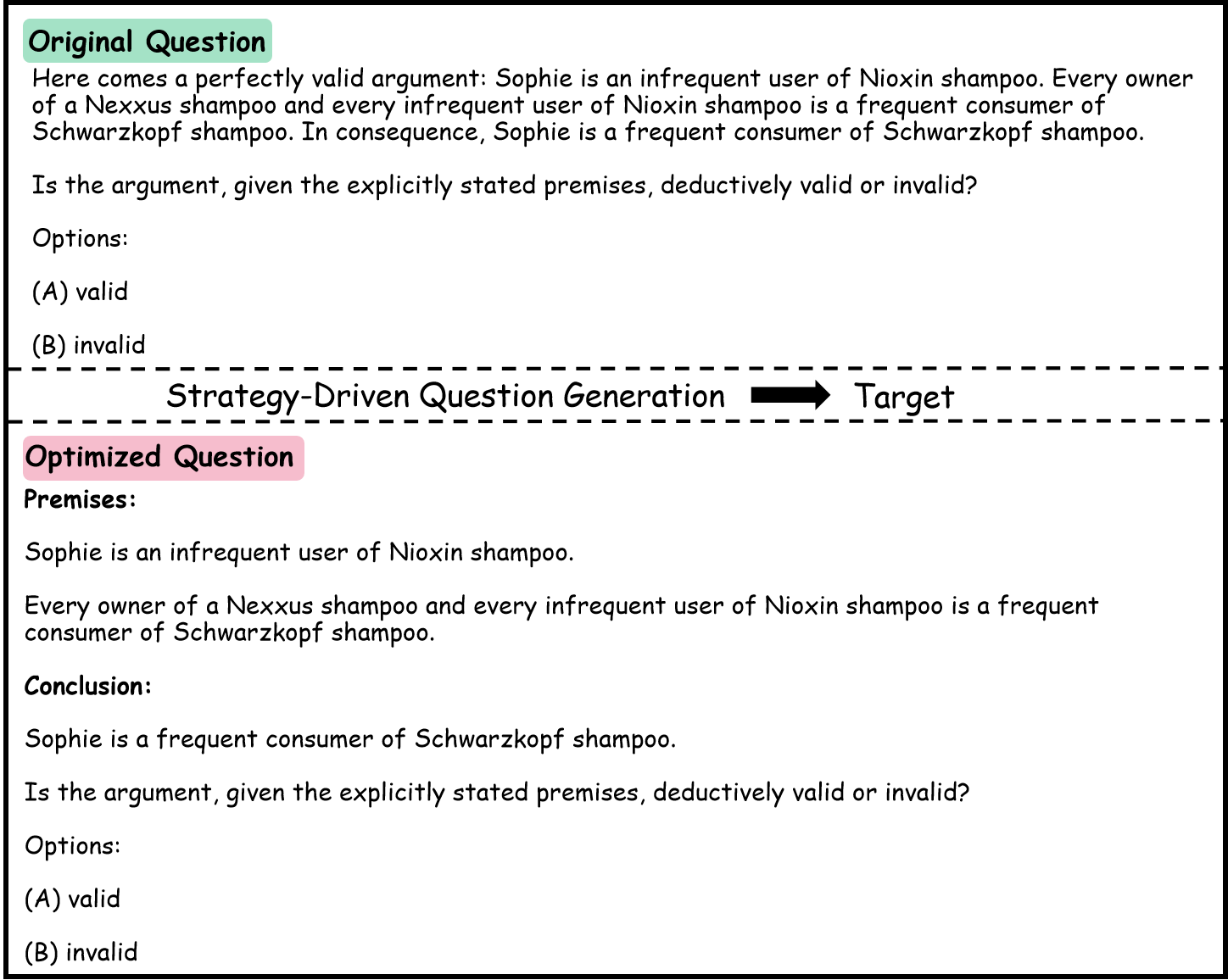}

\caption{Inference-stage strategy-driven question generation on the Formal Fallacies task. 
The \textit{Question-Generator} reformulates the original query following the learned strategy, while the -Judge{Question-Judge} validates semantic preservation and structural quality, producing an optimized question for final model inference.}
\label{fig:case_study2}
\end{figure}

\subsection{Interpretable Helix Co-Evolution Trajectory}

Figures~\ref{fig:case_study} and~\ref{fig:case_study2} present a representative optimization trajectory on the Formal Fallacies task, illustrating how Helix performs structured co-evolution during training and systematically applies learned strategies during inference.

Figure~\ref{fig:case_study} visualizes the training-stage dual-helix process. The \textit{Planner} decomposes the task into explicit helix objectives with paired reformulation and prompting goals (e.g., \textit{Preamble Removal} $\leftrightarrow$ \textit{Step-by-Step Reasoning}), yielding a transparent and task-adaptive optimization trajectory. Within each helix, \textit{Prompt-Architect} and \textit{Question-Architect} iteratively propose and critique updates, ensuring that improvements in one dimension remain compatible with the other. This bidirectional refinement produces tightly aligned prompt strategies and question reformulation policies.

Figure~\ref{fig:case_study2} illustrates the inference-stage strategy application. The \textit{Question-Generator} progressively transforms an original query following the learned reformulation strategy, while the \textit{Question-Judge} enforces semantic preservation and structural quality through multi-criteria validation. This verifies how training-stage co-evolution translates into reliable and interpretable improvements on unseen instances.

Overall, the above results highlight Helix’s ability to decompose complex optimization into structured helix objectives, enforce cross-dimensional consistency through dual-helix critique, and transfer learned strategies to inference in a controllable and interpretable manner.
{\footnote{Appendix~\ref{app:generalization} presents generalization experiments across different LLMs. 
Appendix~\ref{app:agent_prompts} includes detailed agent prompts and interaction examples.
Appendix~\ref{app:optimized_outputs} presents detailed optimization outputs for all twelve tasks.}}

\begin{table}[t!]
\centering
\caption{Component ablation on BBH tasks (accuracy \%). $\Delta$ denotes performance drop relative to the full Helix framework.}
\setlength{\tabcolsep}{1.2pt}
\renewcommand{\arraystretch}{1.2}
\small
\begin{tabular}{l|ccccc|c}
\toprule
\textbf{Variation} & \textbf{D.QA} & \textbf{G.S.} & \textbf{F.F.} & \textbf{R.N.} & \textbf{S.U.} & \textbf{Avg} \\
\hline
Full & 73.20 & 73.20 & 94.05 & 93.20 & 92.40 & 85.21 \\
\hline
w/o Planner & 66.00 & 68.40 & 85.20 & 89.80 & 84.60 & 78.80 \\
\quad $\Delta$ & \textcolor{darkgreen}{(-7.20)} & \textcolor{darkgreen}{(-4.80)} & \textcolor{darkgreen}{(-8.85)} & \textcolor{darkgreen}{(-3.40)} & \textcolor{darkgreen}{(-7.80)} & \textcolor{darkgreen}{(-6.41)} \\
w/o P-Arch & 64.80 & 69.20 & 82.60 & 87.80 & 78.50 & 76.58 \\
\quad $\Delta$ & \textcolor{darkgreen}{(-8.40)} & \textcolor{darkgreen}{(-4.00)} & \textcolor{darkgreen}{(-11.45)} & \textcolor{darkgreen}{(-5.40)} & \textcolor{darkgreen}{(-13.90)} & \textcolor{darkgreen}{(-8.63)} \\
w/o Q-Arch & 66.40 & 70.80 & 84.20 & 88.60 & 75.60 & 77.12 \\
\quad $\Delta$ & \textcolor{darkgreen}{(-6.80)} & \textcolor{darkgreen}{(-2.40)} & \textcolor{darkgreen}{(-9.85)} & \textcolor{darkgreen}{(-4.60)} & \textcolor{darkgreen}{(-16.80)} & \textcolor{darkgreen}{(-8.09)} \\
w/o Mediator & 69.60 & 71.20 & 89.40 & 90.60 & 80.45 & 80.25 \\
\quad $\Delta$ & \textcolor{darkgreen}{(-3.60)} & \textcolor{darkgreen}{(-2.00)} & \textcolor{darkgreen}{(-4.65)} & \textcolor{darkgreen}{(-2.60)} & \textcolor{darkgreen}{(-11.95)} & \textcolor{darkgreen}{(-4.96)} \\
w/o Generator & 70.80 & 72.40 & 90.80 & 91.20 & 90.40 & 77.75 \\
\quad $\Delta$ & \textcolor{darkgreen}{(-2.40)} & \textcolor{darkgreen}{(-0.80)} & \textcolor{darkgreen}{(-3.25)} & \textcolor{darkgreen}{(-2.00)} & \textcolor{darkgreen}{(-2.00)} & \textcolor{darkgreen}{(-7.46)} \\
w/o Judge & 71.50 & 72.80 & 92.00 & 92.00 & 91.70 & 82.00 \\
\quad $\Delta$ & \textcolor{darkgreen}{(-1.70)} & \textcolor{darkgreen}{(-0.40)} & \textcolor{darkgreen}{(-2.05)} & \textcolor{darkgreen}{(-1.20)} & \textcolor{darkgreen}{(-0.70)} & \textcolor{darkgreen}{(-3.21)} \\
\bottomrule
\end{tabular}
\label{tab:component_ablation}
\end{table}

\subsection{Ablation Study}
\label{sec:ablation}

Table~\ref{tab:component_ablation} reports the effect of removing key components on five BBH tasks. 
Eliminating the \textit{Planner} leads to a clear performance drop (6.41\% on average), highlighting the importance of structured task decomposition for effective dual optimization. 
Removing either the \textit{Prompt-Architect} or the \textit{Question-Architect} results in even larger degradations (8.63\% and 8.09\%), as the system collapses to single-dimension optimization, confirming that coordinated refinement across both prompts and questions is essential. 

The \textit{Mediator} contributes substantially to stable co-evolution: without validation, incompatible updates are accepted, yielding a 4.96\% performance decline. 
At inference time, removing the \textit{Question-Generator} prevents applying learned reformulation strategies and effectively reduces Helix to prompt-only optimization, causing significant degradation. 
Finally, removing the \textit{Question-Judge} weakens quality control, allowing suboptimal reformulations to pass and leading to consistent accuracy loss.

Overall, each component plays a complementary role in enabling effective dual-helix co-evolution.

\section{Conclusion}

We propose Helix, a dual-helix co-evolutionary multi-agent framework that integrates Planner-guided decomposition, bidirectional co-evolution, and strategy-driven question generation for joint optimization of questions and prompts. 
This structured design enables efficient and synergistic improvements across both dimensions with strong interpretability and task adaptability. 
Our extensive experimental results demonstrate that prompt optimization benefits from being treated as a coupled interaction design problem.

\newpage

\bibliographystyle{icml2026}
\bibliography{Helix}

\clearpage
\newpage

\appendix

\section{Dataset Details}
\label{app:datasets}

To comprehensively evaluate the dual optimization capabilities of our framework, we select 12 tasks from diverse benchmarks spanning reasoning and knowledge domains.

\paragraph{General Task Evaluation.} We select five tasks from BBH~\cite{suzgun2023challenging}, which consist of challenging reasoning tasks that assess logical inference and problem-solving skills, including Disambiguation QA, Geometric Shapes, Formal Fallacies, Ruin Names, and Sports Understanding. From MMLU~\cite{hendrycks2020measuring}, we select three subject-specific tasks designed to evaluate general knowledge: College Biology, Electrical Engineering, and Marketing. From MMLU-Pro~\cite{wang2024mmlu}, we include History and Philosophy, which feature advanced knowledge questions with more challenging distractors.

\paragraph{Domain-Specific Task Evaluation.} We include two benchmarks: AGIEval~\cite{zhong2024agieval} and AQuA-RAT~\cite{ling2017program}. From AGIEval, a human-centric benchmark for evaluating foundation models, we select the analytical reasoning section of the Law School Admission Test (LSAT-AR). AQuA-RAT is a widely used mathematical reasoning dataset containing algebraic word problems.

\paragraph{Dataset Split.} We adopt a minimal training paradigm by selecting only a single instance from each dataset for optimization. Despite this extremely limited supervision, our method demonstrates strong and consistent performance across diverse datasets. The detailed partition of the dataset is presented in Table~\ref{tab:dataset_split}.

\begin{table}[h]
\centering
\small
\setlength{\tabcolsep}{4pt}
\renewcommand{\arraystretch}{1.15}
\begin{tabular}{llcc}
\toprule
\textbf{Tasks} & \textbf{ABBR.} & \textbf{Train} & \textbf{Test} \\
\midrule
\multicolumn{4}{l}{\textbf{Bigbench}} \\
\quad Disambiguation QA & D.QA & 1 & 249 \\
\quad Geometric Shapes & G.S. & 1 & 249 \\
\quad Formal Fallacies & F.F. & 1 & 249 \\
\quad Ruin Names & R.N. & 1 & 249 \\
\quad Sports Understanding & S.U. & 1 & 249 \\
\midrule
\multicolumn{4}{l}{\textbf{MMLU}} \\
\quad College Biology & C.B. & 1 & 143 \\
\quad Electrical Engineering & E.E. & 1 & 144 \\
\quad Marketing & M.T. & 1 & 233 \\
\midrule
\multicolumn{4}{l}{\textbf{MMLU-Pro}} \\
\quad History & Hist & 1 & 380 \\
\quad Philosophy & Phil & 1 & 498 \\
\midrule
\multicolumn{4}{l}{\textbf{AGIEval}} \\
\quad LSAT-AR & L.A. & 1 & 229 \\
\midrule
\textbf{AQuA-RAT} & A.R. & 1 & 253 \\
\bottomrule
\end{tabular}
\caption{Data split of tasks. One instance for training, remaining instances for testing.}
\label{tab:dataset_split}
 \vspace{-2em}
\end{table}

\section{Experimental Settings and Baselines}
\label{app:settings}

We compare Helix with two categories of baselines: manual prompting strategies and automated prompt optimization (APO) methods.

\paragraph{Manual Prompting Strategies.}
(1) \textbf{No Prompt}: Direct question-answering without any prompt guidance. 
(2) \textbf{CoT}~\cite{wei2022chain}: Chain-of-thought prompting to guide step-by-step reasoning.

\paragraph{Automated Prompt Optimization Methods.}
(3) \textbf{APE}~\cite{zhou2022large}: Generates multiple candidate prompts and performs search-based optimization. 
(4) \textbf{OPRO}~\cite{yang2023large}: Optimizes prompts using sophisticated meta-prompts. 
(5) \textbf{PE2}~\cite{ye2024prompt}: Employs prompt expansion techniques for iterative refinement. 
(6) \textbf{MARS}~\cite{zhang2025mars}: State-of-the-art multi-agent method with Teacher--Critic--Student architecture and Socratic guidance.

\paragraph{Implementation.}
All Helix agents are implemented using GPT-4o~\cite{achiam2023gpt}. GPT-4o exhibits strong reasoning and generation capabilities, efficiently exploring the dual optimization space. 

For the target LLM, we use GPT-4o on BBH tasks and Qwen2.5-32B-Instruct~\cite{qwen2025qwen25technicalreport} on the remaining benchmarks. Qwen2.5-32B-Instruct is a large-scale instruction-tuned language model that demonstrates strong performance across diverse reasoning and knowledge-intensive tasks, making it well-suited for evaluating the generalization of our optimized question-prompt pairs.

We set the number of optimization runs to $T=10$, with at most $R_{\max}=3$ dual-helix co-evolution rounds per helix objective and $K_{\max}=3$ generator--judge iterations during inference. We report the best-performing configuration across all runs.

\section{Generalization Across Different LLM Configurations}
\label{app:generalization}

We validate Helix's generalization capability by testing with different LLMs for Helix agents and target LLMs for task evaluation.

\subsection{Generalization Across Different LLMs for Helix Agents}

We conduct experiments using Gemini models to implement Helix agents. Specifically, we use Gemini-2.5-Pro~\cite{comanici2025gemini} for all Helix agents (\textit{Planner}, \textit{Prompt-Architect}, \textit{Question-Architect}, \textit{Question-Generator}, \textit{Question-Judge}, and \textit{Mediator}), while using Gemini-2.0-Flash as the target LLM.

As shown in Table~\ref{tab:optimizer_llm}, Helix achieves strong performance across all benchmark categories when implemented with Gemini models. The results demonstrate that Helix's dual optimization mechanism is model-agnostic, achieving consistent improvements whether the framework agents are implemented with GPT-4o or Gemini-2.5-Pro.

\begin{table}[h]
\centering
\setlength{\tabcolsep}{2pt}
\renewcommand{\arraystretch}{1.1}
\small
\begin{tabular}{lccccc}
\toprule
\textbf{Models} & \textbf{BBH} & \textbf{AGIEval} & \textbf{MMLU$^*$\footnotemark} & \textbf{AQuA-RAT} & \textbf{Avg} \\
\midrule
No Prompt & 70.16 & 45.65 & 69.72 & 86.61 & 68.04 \\
CoT & 71.36 & 45.22 & 72.68 & 87.40 & 69.17 \\
APE & 72.45 & 43.44 & 71.21 & 85.92 & 68.26 \\
OPRO & 74.12 & 44.35 & 73.40 & 86.78 & 69.66 \\
PE2 & 75.53 & 46.78 & 74.40 & 88.23 & 71.24 \\
MARS & 77.22 & 45.33 & 74.27 & 89.15 & 71.49 \\
\midrule
Q + P-Opt & 78.34 & 47.39 & 75.70 & 90.34 & 72.94 \\
Q-Opt & 73.84 & 46.09 & 73.74 & 87.86 & 70.38 \\
Q-Opt + CoT & 76.16 & 46.96 & 74.75 & 89.67 & 71.89 \\
\midrule
\textbf{Q-Opt + P-Opt} & \textbf{80.11} & \textbf{48.88} & \textbf{77.95} & \textbf{91.73} & \textbf{74.67} \\
\bottomrule
\end{tabular}
\caption{Performance comparison when using Gemini-2.5-Pro for all Helix agents, with Gemini-2.0-Flash as target LLM (accuracy \%).}
\label{tab:optimizer_llm}
 \vspace{-1.5em}
\end{table}
\footnotetext{MMLU$^*$ denotes average performance across MMLU and MMLU-Pro tasks.}

\subsection{Generalization Across Target LLMs}

To evaluate the transferability of optimized question-prompt pairs across different target LLMs, we test Helix outputs using various target LLMs for final evaluation, including GPT-3.5, GPT-4, and GPT-4o.

As shown in Table~\ref{tab:eval_models}, the optimized question-prompt pairs demonstrate strong generalization ability across different target LLMs. Helix consistently outperforms baselines regardless of which target LLM is used for final evaluation.

\begin{table}[h]
\centering
\setlength{\tabcolsep}{3pt}
\renewcommand{\arraystretch}{1.1}
\small
\begin{tabular}{lcccc}
\toprule
\textbf{Method} & \textbf{GPT-3.5} & \textbf{GPT-4} & \textbf{GPT-4o} & \textbf{Avg} \\
\midrule
No Prompt & 82.64 & 84.72 & 89.58 & 85.65 \\
CoT & 85.42 & 87.85 & 92.71 & 88.66 \\
MARS & 86.11 & 88.54 & 93.40 & 89.35 \\
Helix & \textbf{87.85} & \textbf{90.28} & \textbf{95.14} & \textbf{91.09} \\
\bottomrule
\end{tabular}
\caption{Performance comparison on MMLU tasks across different target LLMs (accuracy \%). All Helix agents were implemented using GPT-4o. Results are averaged across C.B., E.E., and M.T. tasks.}
\label{tab:eval_models}
\vspace{-1cm}
\end{table}

\section{Agent Interaction Examples}
\label{app:agent_prompts}

\subsection{Agent Prompts}

The Helix framework employs six specialized agents to accomplish dual optimization. Table~\ref{tab:agent_prompts_detailed} summarizes the prompts for all agents, exemplified with the Formal Fallacies task from the BBH dataset.

In the training stage, \textit{Planner} first performs task decomposition into sequential helix objectives. Then, \textit{Prompt-Architect} and \textit{Question-Architect} engage in bidirectional critique through dual-helix co-evolution, with \textit{Mediator} validating improvements to ensure synergistic compatibility. In the inference stage, \textit{Question-Generator} applies the learned strategy to transform test instances, while \textit{Question-Judge} performs multi-dimensional validation to ensure quality and semantic fidelity. Finally, the target LLM evaluates the optimized question-prompt pair on the test set.

\begin{table*}[t]
\centering
\caption{Summary of agent prompts in the Helix framework. The examples are from the Formal Fallacies task of the BBH dataset.}
\label{tab:agent_prompts_detailed}
\small
\begin{tabular}{p{16.5cm}}
\toprule
\textbf{Input} 

Task Type: Formal Fallacies

Task Description: Evaluate arguments to determine if they are deductively valid or invalid. Problems present arguments with multiple premises and a conclusion, requiring extraction of logical components, tracking of negations and logical operators, evaluation of whether the conclusion necessarily follows from the premises, and binary classification (valid/invalid).

Expected Output Format: A letter answer in parentheses (e.g., "(A)" for valid or "(B)" for invalid)

I want to optimize both the question presentation and the solving prompt to achieve the highest correctness rate on this type of problem. \\
\midrule
\textbf{\textit{Planner}}

Split the task into detailed helix objectives for co-evolution of question presentation and prompt instructions.

For example, for the Disambiguation QA task, the optimization is planned as follows: Total helices: 4. helix 1: Pronoun Highlighting $\leftrightarrow$ Pronoun Identification, where the question goal is to visually highlight the pronoun to identify, the prompt goal is to direct immediate pronoun identification without search, and the connection is that highlighted pronouns enable prompts to skip search steps. helix 2: Antecedent Clarity $\leftrightarrow$ Antecedent Filtering, where the question goal is to present potential antecedents clearly in context, the prompt goal is to guide systematic filtering based on grammatical agreement, and the connection is that clear noun visibility supports efficient antecedent filtering... \\
\midrule
\textbf{\textit{Prompt-Architect}}

Prompt-Design: You are a prompt designer in the Helix framework. Your task is to design an improved prompt that achieves the prompt goal and works with the question strategy...

Question-Critique: You are a question strategy critic in the Helix framework. Your task is to evaluate proposed question strategy to ensure it supports effective prompting...\\
\midrule
\textbf{\textit{Question-Architect}}

Question-Design: You are a question designer in the Helix framework. Your task is to design an improved question strategy that achieves the question goal and works with the prompt...

Prompt-Critique: You are a prompt critic in the Helix framework. Your task is to evaluate proposed prompt to ensure it works effectively with the question strategy... \\
\midrule
\textbf{\textit{Question-Generator}}

You are a question modifier in the Helix framework. Your task is to apply a presentation strategy to an original question, creating a modified version that is clearer while preserving all original content...\\
\midrule
\textbf{\textit{Question-Judge}}

You are a question quality judge in the Helix framework. Your task is to verify that a modified question follows the question strategy and preserves the original meaning... \\
\midrule
\textbf{\textit{Mediator}}

You are the mediator in the Helix framework. Your task is to judge if the optimized prompt satisfies the prompt goal, if their connection works synergistically as planned... \\
\midrule
\textbf{Target LLM}

Optimized Prompt: Analyze the argument step by step to determine its deductive validity. The premises and conclusion are clearly labeled above. Carefully examine the logical structure, paying special attention to negation terms and their logical impact, and assess whether the premises logically necessitate the conclusion. Apply formal logical principles to determine if the argument is valid or invalid.

Optimized Question:

\textbf{Premises:} Sophie is an infrequent user of Nioxin shampoo. Every owner of a Nexxus shampoo and every infrequent user of Nioxin shampoo is a frequent consumer of Schwarzkopf shampoo. 

\textbf{Conclusion:} Sophie is a frequent consumer of Schwarzkopf shampoo. Is the argument, given the explicitly stated premises, deductively valid or invalid?

Options: (A) valid (B) invalid \\
\bottomrule
\end{tabular}
\end{table*}

\subsection{Training Stage}

\begin{figure*}[t]
\centering
\includegraphics[width=\textwidth]{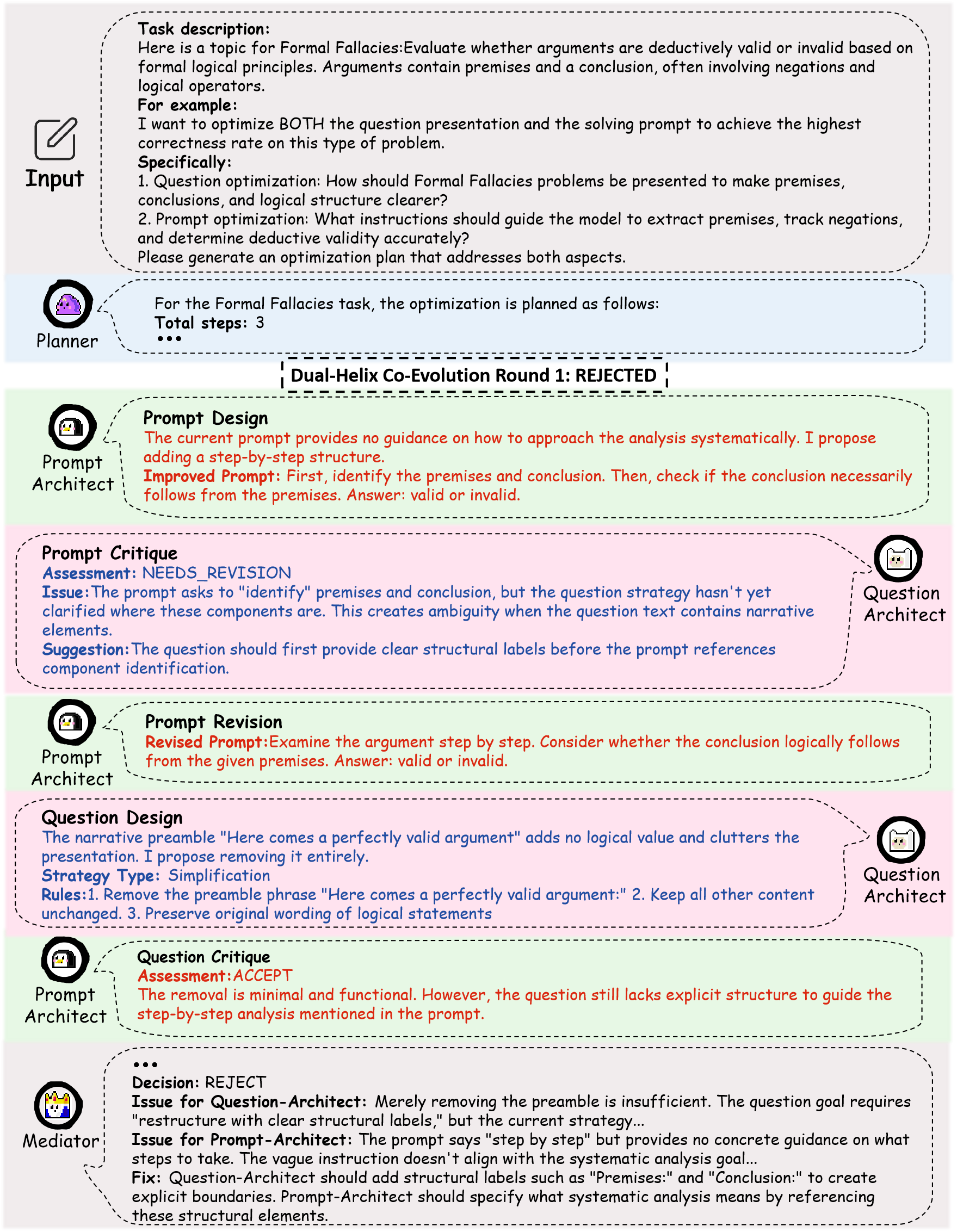}
\caption{Training process of Helix (Part 1/2): Planner-guided decomposition and dual-helix co-evolution Round 1. (Continued in Figure~\ref{fig:sce_process_part2})}
\label{fig:sce_process_part1}
\end{figure*}

\begin{figure*}[t]
\centering
\includegraphics[width=\textwidth]{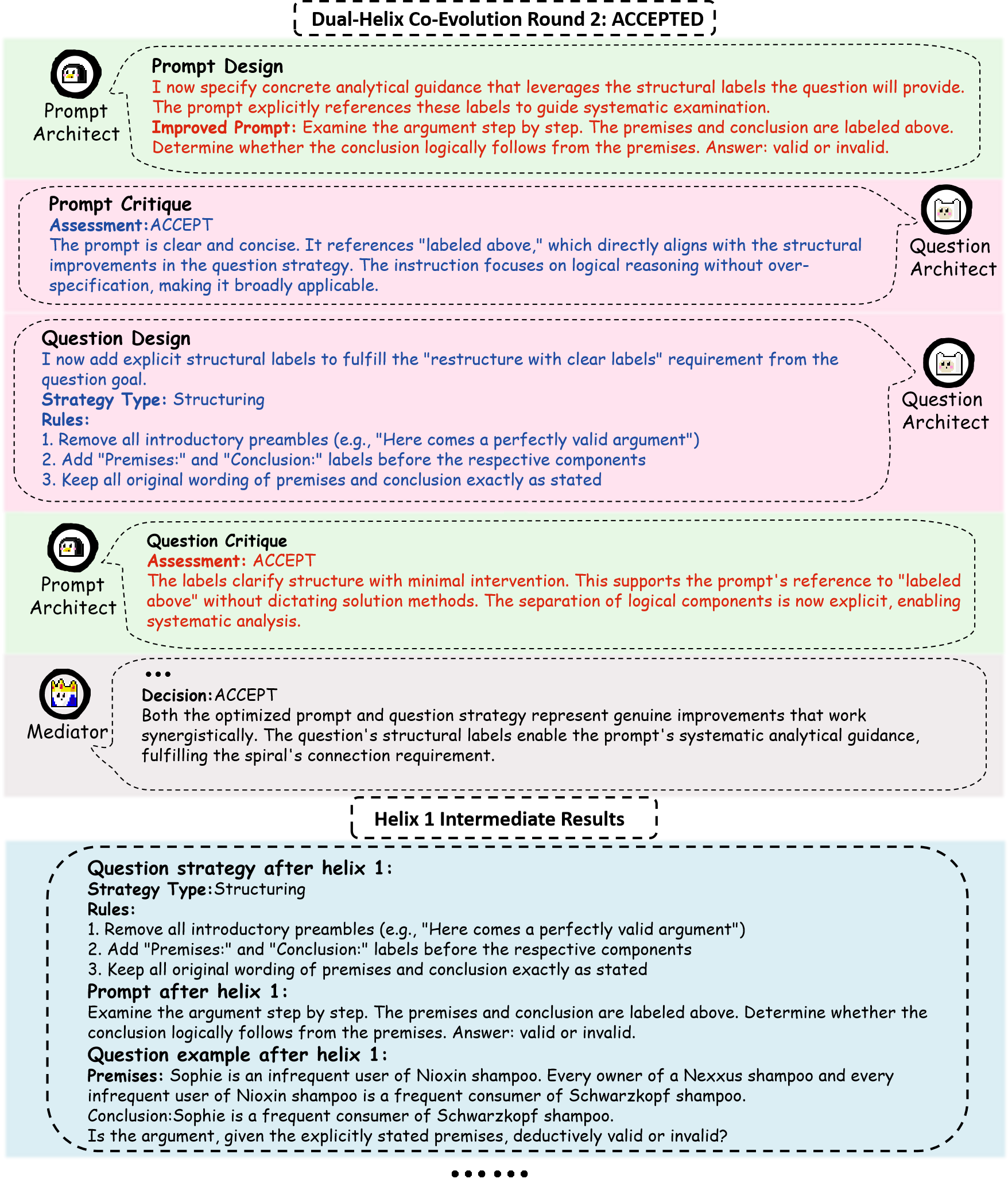}
\caption{Training process of Helix (Part 2/2): Dual-helix co-evolution Round 2.}
\label{fig:sce_process_part2}
\end{figure*}

Figures~\ref{fig:sce_process_part1} and~\ref{fig:sce_process_part2} illustrate the complete training process on the Formal Fallacies task. As shown in Figure~\ref{fig:sce_process_part1}, the training stage begins with planner-guided decomposition, where \textit{Planner} decomposes the task into sequential helix objectives. This is followed by the dual-helix co-evolution process, where \textit{Prompt-Architect} first designs an improved prompt which is then critiqued by \textit{Question-Architect}, followed by \textit{Question-Architect} designing an improved question strategy which is critiqued by \textit{Prompt-Architect}. This bidirectional engagement creates evolutionary pressure toward solutions where both components work synergistically together, enabling the co-evolutionary refinement that distinguishes Helix from single-dimension optimization approaches.

\subsection{Inference Stage}

\begin{figure*}[t]
\centering
\includegraphics[width=\textwidth]{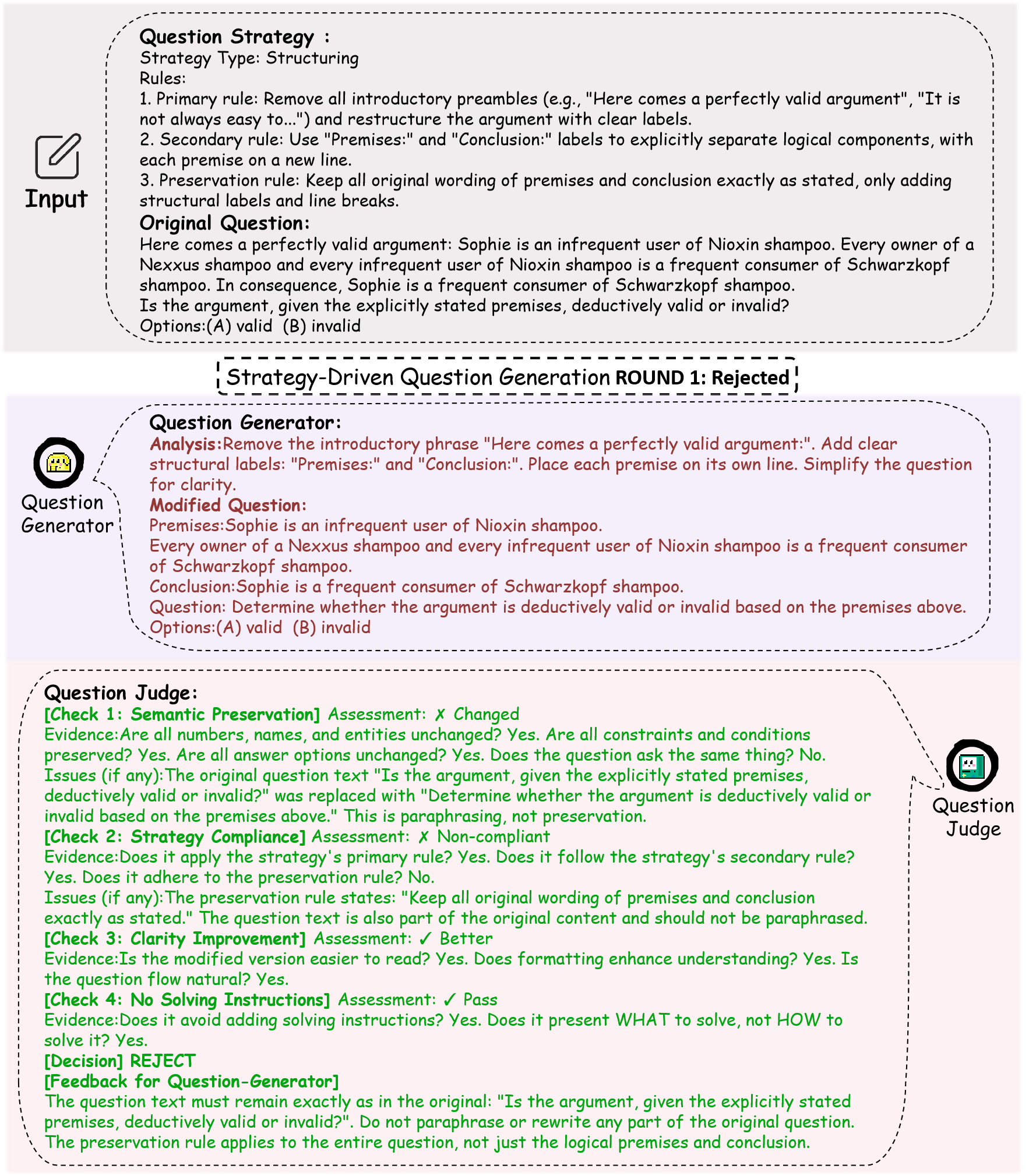}
\caption{Inference process of Helix (Part 1/2): Strategy-driven question generation Round 1. (Continued in Figure~\ref{fig:inference_process_part2})}
\label{fig:inference_process_part1}
\end{figure*}

\begin{figure*}[t]
\centering
\includegraphics[width=\textwidth]{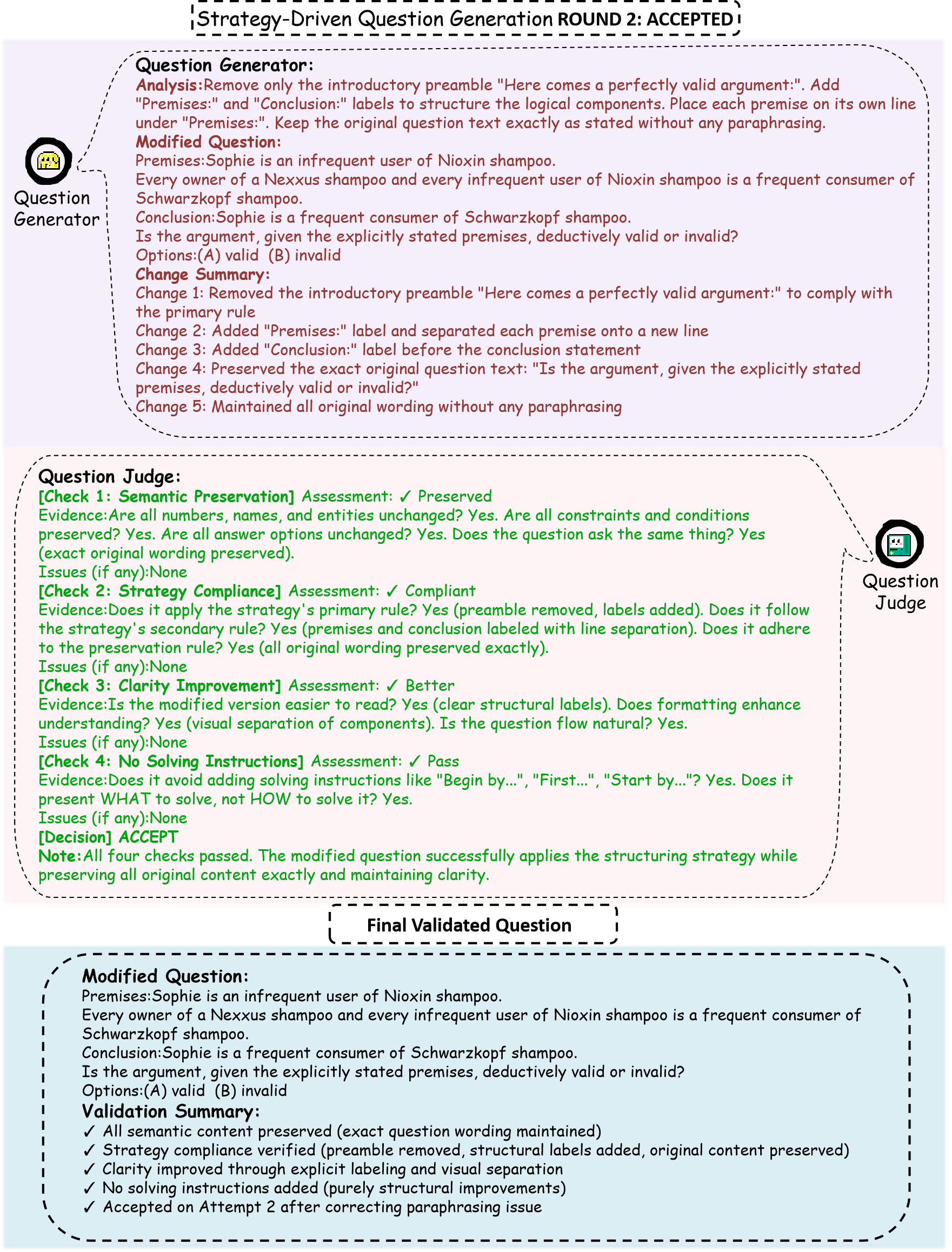}
\caption{Inference process of Helix (Part 2/2): Strategy-driven question generation Round 2.}
\label{fig:inference_process_part2}
\end{figure*}

Figures~\ref{fig:inference_process_part1} and~\ref{fig:inference_process_part2} demonstrate the inference stage where \textit{Question-Generator} applies the learned strategy to transform test instances, with \textit{Question-Judge} performing multi-dimensional validation to ensure semantic preservation and quality.

\section{Detailed Optimization Outputs}
\label{app:optimized_outputs}

This section presents the optimized prompts and question strategies for all twelve tasks, demonstrating the diversity of Helix's optimization outcomes across different reasoning domains. For each task (Disambiguation QA, Geometric Shapes, Formal Fallacies, Ruin Names, Sports Understanding, LSAT-AR, College Biology, Electrical Engineering, Marketing, History, Philosophy, and AQuA-RAT), we present: (1) \textbf{Optimized Prompt} generated by Helix through dual-helix co-evolution, (2) \textbf{Optimized Question Strategy} with transformation rules and strategy type, (3) \textbf{Original Question} from the dataset before optimization, and (4) \textbf{Optimized Question} with the strategy applied. Tables~\ref{tab:dqa_prompt} through Table~\ref{tab:aqua_optimized} sequentially present these optimization results, demonstrating the adaptability and effectiveness of Helix across diverse reasoning and knowledge tasks.

\clearpage


\begin{table*}[t!]
\vspace{-3.0cm}
\caption{Optimized prompt for the Disambiguation QA task generated by Helix through dual-helix co-evolution.}
\vspace{-0.3cm}
\begin{tcolorbox}[
    colback=blue!5!white, 
    colframe=blue!75!black, 
    width=\textwidth, 
    boxrule=1.5pt,
    arc=2mm,
    left=10pt, right=10pt, top=8pt, bottom=8pt
]
\small
Read the sentence carefully. First, locate the highlighted pronoun \texttt{[their]} in the sentence and identify all potential antecedents (nouns) it could refer to. Evaluate each antecedent systematically for grammatical agreement and contextual plausibility. Explicitly eliminate improbable antecedents based on your analysis. If multiple plausible antecedents remain, state that it is ambiguous. Compare your reasoning to the provided options and choose the one that matches your conclusion. Solve step by step.
\end{tcolorbox}
\vspace{-8.0cm}
\label{tab:dqa_prompt}
\end{table*}

\begin{table*}[h]
\caption{Optimized question strategy for the Disambiguation QA task generated by Helix through dual-helix co-evolution.}
\vspace{-0.3cm}
\begin{tcolorbox}[
    colback=green!5!white, 
    colframe=green!75!black, 
    width=\textwidth, 
    boxrule=1.5pt,
    arc=2mm,
    left=10pt, right=10pt, top=8pt, bottom=8pt
]
\small
\textbf{Strategy Type:} Highlighting

\textbf{Rules:}
\begin{enumerate}[leftmargin=*, itemsep=4pt]
    \item \textbf{Primary rule:} Enclose key pronouns (e.g., \texttt{[their]}) in brackets within the provided sentence to ensure they are visually distinct and immediately identifiable.
    \item \textbf{Secondary rule:} Apply the brackets uniformly to all instances of the pronoun in the sentence and related options to maintain consistency.
    \item \textbf{Preservation rule:} Avoid altering the original sentence structure, phrasing, or answer options to preserve the natural flow of the question.
\end{enumerate}
\end{tcolorbox}
\vspace{-8.0cm}
\end{table*}

\begin{table*}[h]
\caption{Original question from the Disambiguation QA task before optimization.}
\vspace{-0.3cm}
\begin{tcolorbox}[
    colback=gray!10!white, 
    colframe=gray!60!black, 
    width=\textwidth, 
    boxrule=1.5pt,
    arc=2mm,
    left=10pt, right=10pt, top=8pt, bottom=8pt
]
\small
In the following sentences, explain the antecedent of the pronoun (which thing the pronoun refers to), or state that it is ambiguous.

Sentence: My parent called their secretary for more information.

Options:

(A) They were my parent's secretary

(B) They were the secretary's secretary

(C) Ambiguous
\end{tcolorbox}
\vspace{-8.0cm}
\end{table*}

\begin{table*}[h]
\caption{Optimized question for the Disambiguation QA task with the strategy applied.}
\vspace{-0.3cm}
\begin{tcolorbox}[
    colback=gray!5!white, 
    colframe=gray!75!black, 
    width=\textwidth, 
    boxrule=1.5pt,
    arc=2mm,
    left=10pt, right=10pt, top=8pt, bottom=8pt
]
\small
In the following sentences, explain the antecedent of the pronoun (which thing the pronoun refers to), or state that it is ambiguous.

Sentence: My parent called \texttt{[their]} secretary for more information.

Options:

(A) \texttt{[They]} were my parent's secretary

(B) \texttt{[They]} were the secretary's secretary

(C) Ambiguous
\end{tcolorbox}
\end{table*}

\clearpage


\begin{table*}[t!]
\caption{Optimized prompt for the Geometric Shapes task generated by Helix through dual-helix co-evolution.}
\vspace{-0.3cm}
\begin{tcolorbox}[
    colback=blue!5!white, 
    colframe=blue!75!black, 
    width=\textwidth, 
    boxrule=1.5pt,
    arc=2mm,
    left=10pt, right=10pt, top=8pt, bottom=8pt
]
\small
First, break down each SVG path command and its coordinates to clearly understand their functions. Ensure all coordinate pairs are unique by checking for duplicates. Identify any ambiguous or incomplete path commands and resolve them by referring to standard SVG path syntax. Count the number of vertices or segments formed by these commands. Categorize potential shapes by the number of vertices or segments, from simplest (e.g., line, triangle) to most complex (e.g., octagon, heptagon). Use this categorization to match your findings to the shape options provided. Present your answer as: Answer: (X)
\end{tcolorbox}
\vspace{-2.5cm}
\end{table*}

\begin{table*}[h]
\caption{Optimized question strategy for the Geometric Shapes task generated by Helix through dual-helix co-evolution.}
\vspace{-0.3cm}
\begin{tcolorbox}[
    colback=green!5!white, 
    colframe=green!75!black, 
    width=\textwidth, 
    boxrule=1.5pt,
    arc=2mm,
    left=10pt, right=10pt, top=8pt, bottom=8pt
]
\small
\textbf{Strategy Type:} Formatting

\textbf{Rules:}
\begin{enumerate}[leftmargin=*, itemsep=4pt]
    \item \textbf{Primary rule:} Include a concise clarification in the question explicitly stating whether the path is open, closed, or subject to specific rules about near-collinear points, overlapping paths, or self-intersections.
    \item \textbf{Secondary rule:} Add this clarification as a note in parentheses or as a short sentence at the end of the question, ensuring it integrates naturally without disrupting readability or overloading the question with excessive detail.
    \item \textbf{Preservation rule:} Keep the original structure, phrasing, and answer options intact to maintain the natural flow and intent of the question.
\end{enumerate}
\end{tcolorbox}
\vspace{-2.5cm}
\end{table*}

\begin{table*}[h]
\caption{Original question from the Geometric Shapes task before optimization.}
\vspace{-0.3cm}
\begin{tcolorbox}[
    colback=gray!10!white, 
    colframe=gray!60!black, 
    width=\textwidth, 
    boxrule=1.5pt,
    arc=2mm,
    left=10pt, right=10pt, top=8pt, bottom=8pt
]
\small
This SVG path element \texttt{<path d="M 25.00,38.00 L 89.00,58.00"/>} draws a shape.

Options:

(A) circle

(B) heptagon

(C) hexagon

(D) kite

(E) line

(F) octagon

(G) pentagon

(H) rectangle

(I) sector

(J) triangle
\end{tcolorbox}
\vspace{-2.5cm}
\end{table*}

\begin{table*}[h]
\caption{Optimized question for the Geometric Shapes task with the strategy applied.}
\vspace{-0.3cm}
\begin{tcolorbox}[
    colback=gray!5!white, 
    colframe=gray!75!black, 
    width=\textwidth, 
    boxrule=1.5pt,
    arc=2mm,
    left=10pt, right=10pt, top=8pt, bottom=8pt
]
\small
This SVG path element \texttt{<path d="M 25.00,38.00 L 89.00,58.00"/>} draws a shape. \textbf{(Note: The path described is an open path, as it does not return to its starting point.)}

Options:

(A) circle

(B) heptagon

(C) hexagon

(D) kite

(E) line

(F) octagon

(G) pentagon

(H) rectangle

(I) sector

(J) triangle
\end{tcolorbox}
\end{table*}

\clearpage


\begin{table*}[t!]
\vspace{-3.0cm}
\caption{Optimized prompt for the Formal Fallacies task generated by Helix through dual-helix co-evolution.}
\vspace{-0.3cm}
\begin{tcolorbox}[
    colback=blue!5!white, 
    colframe=blue!75!black, 
    width=\textwidth, 
    boxrule=1.5pt,
    arc=2mm,
    left=10pt, right=10pt, top=8pt, bottom=8pt
]
\small
Analyze the argument step by step to determine its deductive validity. The premises and conclusion are clearly labeled above. Carefully examine the logical structure, paying special attention to negation terms and their logical impact, and assess whether the premises logically necessitate the conclusion. Apply formal logical principles to determine if the argument is valid or invalid.
\end{tcolorbox}
\vspace{-8.0cm}
\end{table*}

\begin{table*}[h]
\caption{Optimized question strategy for the Formal Fallacies task generated by Helix through dual-helix co-evolution.}
\vspace{-0.3cm}
\begin{tcolorbox}[
    colback=green!5!white, 
    colframe=green!75!black, 
    width=\textwidth, 
    boxrule=1.5pt,
    arc=2mm,
    left=10pt, right=10pt, top=8pt, bottom=8pt
]
\small
\textbf{Strategy Type:} Structuring

\textbf{Rules:}
\begin{enumerate}[leftmargin=*, itemsep=4pt]
    \item \textbf{Primary rule:} Remove all introductory preambles (e.g., ``Here comes a perfectly valid argument'', ``It is not always easy to...'') and restructure the argument with clear labels.
    \item \textbf{Secondary rule:} Use ``\textbf{Premises:}'' and ``\textbf{Conclusion:}'' labels to explicitly separate logical components, with each premise on a new line.
    \item \textbf{Preservation rule:} Keep all original wording of premises and conclusion exactly as stated, only adding structural labels and line breaks.
\end{enumerate}
\end{tcolorbox}
\vspace{-8.0cm}
\end{table*}

\begin{table*}[h]
\caption{Original question from the Formal Fallacies task before optimization.}
\vspace{-0.3cm}
\begin{tcolorbox}[
    colback=gray!10!white, 
    colframe=gray!60!black, 
    width=\textwidth, 
    boxrule=1.5pt,
    arc=2mm,
    left=10pt, right=10pt, top=8pt, bottom=8pt
]
\small
Here comes a perfectly valid argument: Sophie is an infrequent user of Nioxin shampoo. Every owner of a Nexxus shampoo and every infrequent user of Nioxin shampoo is a frequent consumer of Schwarzkopf shampoo. In consequence, Sophie is a frequent consumer of Schwarzkopf shampoo.

Is the argument, given the explicitly stated premises, deductively valid or invalid?

Options:

(A) valid

(B) invalid
\end{tcolorbox}
\vspace{-8.0cm}
\end{table*}

\begin{table*}[h]
\caption{Optimized question for the Formal Fallacies task with the strategy applied.}
\vspace{-0.3cm}
\begin{tcolorbox}[
    colback=gray!5!white, 
    colframe=gray!75!black, 
    width=\textwidth, 
    boxrule=1.5pt,
    arc=2mm,
    left=10pt, right=10pt, top=8pt, bottom=8pt
]
\small
\textbf{Premises:}

Sophie is an infrequent user of Nioxin shampoo.

Every owner of a Nexxus shampoo and every infrequent user of Nioxin shampoo is a frequent consumer of Schwarzkopf shampoo.

\textbf{Conclusion:}

Sophie is a frequent consumer of Schwarzkopf shampoo.

Is the argument, given the explicitly stated premises, deductively valid or invalid?

Options:

(A) valid

(B) invalid
\end{tcolorbox}
\end{table*}

\clearpage


\begin{table*}[t!]
\vspace{-2.0cm}
\caption{Optimized prompt for the Ruin Names task generated by Helix through dual-helix co-evolution.}
\vspace{-0.3cm}
\begin{tcolorbox}[
    colback=blue!5!white, 
    colframe=blue!75!black, 
    width=\textwidth, 
    boxrule=1.5pt,
    arc=2mm,
    left=10pt, right=10pt, top=8pt, bottom=8pt
]
\small
Given an artist, band, or movie name, identify which single-character edit creates the most humorous result. The edit must involve only one character change (adding, removing, or substituting one letter) and should prioritize simplicity, surprise, and absurdity to evoke humor effectively. Ensure the edit maintains a clear connection to the original name, making the humor immediately recognizable and universally understandable, while avoiding overly specific or niche references.

\textbf{Key Guidelines:} (1) Simplicity and Surprise, (2) Cultural Universality, (3) Absurdity and Creativity

\textbf{Evaluation Process:} First, identify which options create recognizable new words versus random character changes. Focus your analysis on options that form actual words or meaningful concepts. For each valid edit, evaluate: Original meaning → New meaning (Does this transformation create humor?), Semantic gap (How absurd or unexpected is the shift?), Universal recognition (Is the new word/concept widely understood?), Immediate humor (Would this edit make someone laugh without explanation?). Eliminate options that create meaningless letter combinations, result in minor changes, or rely on niche knowledge.

\textbf{Selection Strategy:} Prioritize edits where the new word is common and universally known, the semantic transformation is simple and surprising, and the connection to the original remains obvious.
\end{tcolorbox}
\vspace{-6.0cm}
\end{table*}

\begin{table*}[h]
\caption{Optimized question strategy for the Ruin Names task generated by Helix through dual-helix co-evolution.}
\vspace{-0.3cm}
\begin{tcolorbox}[
    colback=green!5!white, 
    colframe=green!75!black, 
    width=\textwidth, 
    boxrule=1.5pt,
    arc=2mm,
    left=10pt, right=10pt, top=8pt, bottom=8pt
]
\small
\textbf{Strategy Type:} Highlighting

\textbf{Rules:}
\begin{enumerate}[leftmargin=*, itemsep=4pt]
    \item \textbf{Primary rule:} Use brackets [ ] to highlight the changed character in each option (e.g., ``fall out bo\texttt{[t]}'').
    \item \textbf{Secondary rule:} After each option, add a brief hint about the result: ``(new word)'' for meaningful edits or ``(random)'' for meaningless ones.
    \item \textbf{Preservation rule:} Keep all original content unchanged.
\end{enumerate}
\end{tcolorbox}
\vspace{-6.0cm}
\end{table*}

\begin{table*}[h]
\caption{Original question from the Ruin Names task before optimization.}
\vspace{-0.3cm}
\begin{tcolorbox}[
    colback=gray!10!white, 
    colframe=gray!60!black, 
    width=\textwidth, 
    boxrule=1.5pt,
    arc=2mm,
    left=10pt, right=10pt, top=8pt, bottom=8pt
]
\small
Which of the following is a humorous edit of this artist or movie name: 'fall out boy'?

Options:

(A) faqll out boy

(B) fall out bot

(C) falbl out boy

(D) fall outt boy
\end{tcolorbox}
\vspace{-6.0cm}
\end{table*}

\begin{table*}[h]
\caption{Optimized question for the Ruin Names task with the strategy applied.}
\vspace{-0.3cm}
\begin{tcolorbox}[
    colback=gray!5!white, 
    colframe=gray!75!black, 
    width=\textwidth, 
    boxrule=1.5pt,
    arc=2mm,
    left=10pt, right=10pt, top=8pt, bottom=8pt
]
\small
Which of the following is a humorous edit of this artist or movie name: 'fall out boy'?

Options:

(A) fa\texttt{[q]}ll out boy (random)

(B) fall out bo\texttt{[t]} (new word)

(C) fal\texttt{[b]}l out boy (random)

(D) fall ou\texttt{[tt]} boy (random)
\end{tcolorbox}
\end{table*}

\clearpage


\begin{table*}[t!]
\vspace{-4.0cm}
\caption{Optimized prompt for the Sports Understanding task generated by Helix through dual-helix co-evolution.}
\vspace{-0.3cm}
\begin{tcolorbox}[
    colback=blue!5!white, 
    colframe=blue!75!black, 
    width=\textwidth, 
    boxrule=1.5pt,
    arc=2mm,
    left=10pt, right=10pt, top=8pt, bottom=8pt
]
\small
First, identify the athlete by focusing on their visually distinct name to determine their primary sport. Next, analyze the highlighted action or terminology within the sentence to verify whether it specifically belongs to that sport. Compare the highlighted action to typical actions in the athlete's sport to check for any cross-sport mismatch. Then, consider the event context (e.g., competition type, event setting) and evaluate whether it aligns logically with both the athlete's sport and the described action. For rare or niche same-sport actions, assess plausibility by considering: (1) the athlete's known skill set or expertise, (2) historical precedent for similar actions, and (3) whether the event's competitive level makes the rare action feasible. Answer yes or no.
\end{tcolorbox}
\vspace{-10.0cm}
\end{table*}

\begin{table*}[h]
\caption{Optimized question strategy for the Sports Understanding task generated by Helix through dual-helix co-evolution.}
\vspace{-0.3cm}
\begin{tcolorbox}[
    colback=green!5!white, 
    colframe=green!75!black, 
    width=\textwidth, 
    boxrule=1.5pt,
    arc=2mm,
    left=10pt, right=10pt, top=8pt, bottom=8pt
]
\small
\textbf{Strategy Type:} Highlighting

\textbf{Rules:}
\begin{enumerate}[leftmargin=*, itemsep=4pt]
    \item \textbf{Primary rule:} Use double asterisks around the athlete's name to make it bold and immediately visible (e.g., \textbf{Neymar}).
    \item \textbf{Secondary rule:} Keep all other content exactly as is - do NOT highlight actions, terminology, or event names.
    \item \textbf{Preservation rule:} Keep the original structure, phrasing, and event context intact to ensure the question reads naturally and aligns with real-world plausibility.
\end{enumerate}
\end{tcolorbox}
\vspace{-10.0cm}
\end{table*}

\begin{table*}[h]
\caption{Original question from the Sports Understanding task before optimization.}
\vspace{-0.3cm}
\begin{tcolorbox}[
    colback=gray!10!white, 
    colframe=gray!60!black, 
    width=\textwidth, 
    boxrule=1.5pt,
    arc=2mm,
    left=10pt, right=10pt, top=8pt, bottom=8pt
]
\small
Is the following sentence plausible? "Neymar did a maradona on the defender in the Champions League Semifinal."
\end{tcolorbox}
\vspace{-10.0cm}
\end{table*}

\begin{table*}[h]
\caption{Optimized question for the Sports Understanding task with the strategy applied.}
\vspace{-0.3cm}
\begin{tcolorbox}[
    colback=gray!5!white, 
    colframe=gray!75!black, 
    width=\textwidth, 
    boxrule=1.5pt,
    arc=2mm,
    left=10pt, right=10pt, top=8pt, bottom=8pt
]
\small
Is the following sentence plausible? "\textbf{Neymar} did a maradona on the defender in the Champions League Semifinal."
\end{tcolorbox}
\end{table*}

\clearpage


\begin{table*}[t!]

\caption{Optimized prompt for the LSAT-AR task generated by Helix through dual-helix co-evolution.}
\vspace{-0.3cm}
\begin{tcolorbox}[
    colback=blue!5!white, 
    colframe=blue!75!black, 
    width=\textwidth, 
    boxrule=1.5pt,
    arc=2mm,
    left=10pt, right=10pt, top=8pt, bottom=8pt
]
\small
Carefully analyze the given scheduling problem step by step, prioritizing constraint-based logical reasoning to ensure thorough evaluation.

\textbf{Step 1:} List and classify every constraint in the problem. Identify which constraints specify fixed requirements that lock elements to positions, which forbid certain arrangements, and which create conditional dependencies. Ensure no constraint is overlooked.

\textbf{Step 2:} Apply constraints in order of restrictiveness. Begin with fixed requirements that directly determine positions or immediately eliminate options. Then verify prohibitions are not violated. Finally check that all conditional relationships hold true when their trigger conditions are met.

\textbf{Step 3:} Evaluate each answer option through complete constraint verification. For each option, systematically check it against every single constraint one by one. Mark each constraint as satisfied or violated. An option is valid only if it satisfies all constraints without exception.

\textbf{Step 4:} Before finalizing your answer, perform reverse verification. Take your selected option and re-check it against each constraint to confirm no violations were missed. Do not make assumptions beyond what is explicitly stated in the problem.

By following this methodical approach with complete constraint coverage, you will systematically eliminate all incorrect options and identify the correct answer with maximum accuracy.
\end{tcolorbox}
\vspace{-0.5cm}
\end{table*}

\begin{table*}[h]
\caption{Optimized question strategy for the LSAT-AR task generated by Helix through dual-helix co-evolution.}
\vspace{-0.3cm}
\begin{tcolorbox}[
    colback=green!5!white, 
    colframe=green!75!black, 
    width=\textwidth, 
    boxrule=1.5pt,
    arc=2mm,
    left=10pt, right=10pt, top=8pt, bottom=8pt
]
\small
\textbf{Strategy Type:} Structuring

\textbf{Rules:}
\begin{enumerate}[leftmargin=*, itemsep=4pt]
    \item \textbf{Primary rule:} Group constraints by type using category headers. Format as ``\textbf{Fixed Assignments:}'' followed by constraints, then ``\textbf{Prohibitions:}'' followed by constraints, then ``\textbf{Conditional Rules:}'' followed by constraints.
    \item \textbf{Secondary rule:} Fixed Assignments specify definite requirements (e.g., ``Tuesday is the only day George can report''). Prohibitions forbid arrangements (e.g., ``Neither Olivia nor Robert can give afternoon reports''). Conditional Rules have if-then structure (e.g., ``If Nina gives a report, then Helen and Irving must both give reports the next day'').
    \item \textbf{Preservation rule:} Keep all constraint text verbatim. Do not paraphrase or modify context or options. Only add category headers.
\end{enumerate}
\end{tcolorbox}
\vspace{-0.5cm}
\end{table*}

\begin{table*}[h]
\caption{Original question from the LSAT-AR task before optimization.}
\vspace{-0.3cm}
\begin{tcolorbox}[
    colback=gray!10!white, 
    colframe=gray!60!black, 
    width=\textwidth, 
    boxrule=1.5pt,
    arc=2mm,
    left=10pt, right=10pt, top=8pt, bottom=8pt
]
\small
Of the eight students—George, Helen, Irving, Kyle, Lenore, Nina, Olivia, and Robert—in a seminar, exactly six will give individual oral reports during three consecutive days—Monday, Tuesday, and Wednesday. Exactly two reports will be given each day—one in the morning and one in the afternoon—according to the following conditions: Tuesday is the only day on which George can give a report. Neither Olivia nor Robert can give an afternoon report. If Nina gives a report, then on the next day Helen and Irving must both give reports, unless Nina's report is given on Wednesday.

Which one of the following is a pair of students who, if they give reports on the same day as each other, must give reports on Wednesday?

Options:
...
\end{tcolorbox}
\vspace{-0.5cm}
\end{table*}

\begin{table*}[h]
\caption{Optimized question for the LSAT-AR task with the strategy applied.}
\vspace{-0.3cm}
\begin{tcolorbox}[
    colback=gray!5!white, 
    colframe=gray!75!black, 
    width=\textwidth, 
    boxrule=1.5pt,
    arc=2mm,
    left=10pt, right=10pt, top=8pt, bottom=8pt
]
\small
Of the eight students—George, Helen, Irving, Kyle, Lenore, Nina, Olivia, and Robert—in a seminar, exactly six will give individual oral reports during three consecutive days—Monday, Tuesday, and Wednesday. Exactly two reports will be given each day—one in the morning and one in the afternoon—according to the following conditions:

\textbf{Fixed Assignments:}

\quad • Tuesday is the only day on which George can give a report.

\textbf{Prohibitions:}

\quad • Neither Olivia nor Robert can give an afternoon report.

\textbf{Conditional Rules:}

\quad • If Nina gives a report, then on the next day Helen and Irving must both give reports, unless Nina's report is given on Wednesday.

Which one of the following is a pair of students who, if they give reports on the same day as each other, must give reports on Wednesday?

Options:
...
\end{tcolorbox}
\end{table*}

\clearpage


\begin{table*}[t!]
\vspace{-1.0cm}
\caption{Optimized prompt for the College Biology task generated by Helix through dual-helix co-evolution.}
\vspace{-0.3cm}
\begin{tcolorbox}[
    colback=blue!5!white, 
    colframe=blue!75!black, 
    width=\textwidth, 
    boxrule=1.5pt,
    arc=2mm,
    left=10pt, right=10pt, top=8pt, bottom=8pt
]
\small
Answer college-level biology questions by first identifying the specific biological domain being tested, then applying higher-order thinking skills grounded in established biological principles within that domain.

Begin by carefully reading the \textbf{Domain} label provided at the start of each question. This label indicates which biological field is being tested, such as Cell Biology, Molecular Biology, Ecology \& Population, Genetics \& Evolution, Physiology, or other specialized areas. Use this domain identification to activate your knowledge of the fundamental principles, mechanisms, and reasoning strategies characteristic of that field.

Once you have identified the relevant domain and activated the appropriate knowledge base, apply higher-order thinking skills to analyze the question. Use \textbf{application} to apply established biological principles to novel scenarios. Employ \textbf{analysis} to break down complex biological systems, identify key variables, trace causal pathways, and distinguish between similar but mechanistically distinct processes. Use \textbf{synthesis} to integrate information from multiple biological levels or sub-disciplines.

Systematically evaluate each option using domain-specific biological reasoning. Ground all reasoning in established biological science, basing conclusions on well-established principles and current research findings. Pay close attention to the specific biological context provided. Focus on mechanistic understanding of how and why biological processes occur rather than simple memorization.
\end{tcolorbox}
\vspace{-4.0cm}
\end{table*}

\begin{table*}[h]
\caption{Optimized question strategy for the College Biology task generated by Helix through dual-helix co-evolution.}
\vspace{-0.3cm}
\begin{tcolorbox}[
    colback=green!5!white, 
    colframe=green!75!black, 
    width=\textwidth, 
    boxrule=1.5pt,
    arc=2mm,
    left=10pt, right=10pt, top=8pt, bottom=8pt
]
\small
\textbf{Strategy Type:} Structuring

\textbf{Rules:}
\begin{enumerate}[leftmargin=*, itemsep=4pt]
    \item \textbf{Primary rule:} Add a concise domain label at the beginning of each question to classify the biological field being tested.
    \item \textbf{Format:} Place ``\textbf{Domain: [Field Name]}'' on a separate line before the question.
    \item \textbf{Preservation rule:} Keep ALL original question and option text completely unchanged.
\end{enumerate}
\end{tcolorbox}
\vspace{-4.0cm}
\end{table*}

\begin{table*}[h]
\caption{Original question from the College Biology task before optimization.}
\vspace{-0.3cm}
\begin{tcolorbox}[
    colback=gray!10!white, 
    colframe=gray!60!black, 
    width=\textwidth, 
    boxrule=1.5pt,
    arc=2mm,
    left=10pt, right=10pt, top=8pt, bottom=8pt
]
\small
Based on the characteristic population curves that result from plotting population growth of a species, the most effective means of controlling the mosquito population is to

Options:

(A) maintain the population at a point corresponding to the midpoint of its logistic curve

(B) opt for zero population control once the K value of the curve has been reached

(C) reduce the carrying capacity of the environment to lower the K value

(D) increase the mortality rate
\end{tcolorbox}
\vspace{-4.0cm}
\end{table*}

\begin{table*}[h]
\caption{Optimized question for the College Biology task with the strategy applied.}
\vspace{-0.3cm}
\begin{tcolorbox}[
    colback=gray!5!white, 
    colframe=gray!75!black, 
    width=\textwidth, 
    boxrule=1.5pt,
    arc=2mm,
    left=10pt, right=10pt, top=8pt, bottom=8pt
]
\small
\textbf{Domain: Ecology \& Population}

Based on the characteristic population curves that result from plotting population growth of a species, the most effective means of controlling the mosquito population is to

Options:

(A) maintain the population at a point corresponding to the midpoint of its logistic curve

(B) opt for zero population control once the K value of the curve has been reached

(C) reduce the carrying capacity of the environment to lower the K value

(D) increase the mortality rate
\end{tcolorbox}
\end{table*}

\clearpage


\begin{table*}[t!]
\vspace{-1.0cm}
\caption{Optimized prompt for the Electrical Engineering task generated by Helix through dual-helix co-evolution.}
\vspace{-0.3cm}
\begin{tcolorbox}[
    colback=blue!5!white, 
    colframe=blue!75!black, 
    width=\textwidth, 
    boxrule=1.5pt,
    arc=2mm,
    left=10pt, right=10pt, top=8pt, bottom=8pt
]
\small
To solve electrical engineering problems presented in structured format:

\textbf{1. Extract Information from Structure:} Given parameters are listed with labels and units (e.g., ``Input voltage: 110 V''). Question is stated separately after parameters. Options are grouped clearly. For knowledge questions without parameters, structure remains simple.

\textbf{2. Identify Problem Type:} Knowledge-based (Definitions, properties, standard practices) or Calculation-based (Circuit analysis, formula application).

\textbf{3. For Knowledge Questions:} Focus on technical terms (often highlighted). Recall fundamental EE concepts. Use elimination for uncertain options.

\textbf{4. For Calculation Questions:} Identify circuit type from given parameters. Match to relevant formula. Substitute values (already organized with units). Convert if needed: mA$\rightarrow$A, k$\Omega$$\rightarrow$$\Omega$, $\mu$F$\rightarrow$F.

\textbf{5. Key Formulas:} Ohm's Law (V = IR), Power (P = VI = I²R = V²/R), Duty cycle step-up (D = 1 - V\_in/V\_out), Reactance (X\_C = 1/(2$\pi$fC), X\_L = 2$\pi$fL), Time constant ($\tau$\_RC = RC, $\tau$\_RL = L/R).

Now solve step by step and give the correct answer.
\end{tcolorbox}
\vspace{-3.5cm}
\end{table*}

\begin{table*}[h]
\caption{Optimized question strategy for the Electrical Engineering task generated by Helix through dual-helix co-evolution.}
\vspace{-0.3cm}
\begin{tcolorbox}[
    colback=green!5!white, 
    colframe=green!75!black, 
    width=\textwidth, 
    boxrule=1.5pt,
    arc=2mm,
    left=10pt, right=10pt, top=8pt, bottom=8pt
]
\small
\textbf{Strategy Type:} Structuring

\textbf{Rules:}
\begin{enumerate}[leftmargin=*, itemsep=4pt]
    \item \textbf{Primary:} Separate the question into distinct information blocks - given parameters in one section, the actual question in another section, and options clearly grouped.
    \item \textbf{Secondary:} Highlight numerical values with their units (keep them together as ``110 V'' not ``110'' and ``V'' separately) and emphasize technical terms that are central to the question.
    \item \textbf{Preservation:} Maintain all original numbers, units, technical terms, and option content exactly as given without any modification.
\end{enumerate}
\end{tcolorbox}
\vspace{-3.5cm}
\end{table*}

\begin{table*}[h]
\caption{Original question from the Electrical Engineering task before optimization.}
\vspace{-0.3cm}
\begin{tcolorbox}[
    colback=gray!10!white, 
    colframe=gray!60!black, 
    width=\textwidth, 
    boxrule=1.5pt,
    arc=2mm,
    left=10pt, right=10pt, top=8pt, bottom=8pt
]
\small
SCR gate cathode characteristic is a straight line of 130. Triggered source volume is 15 V. Allowable gate power dissipation is 0.5 W. Compute the gate source resistance.

Options:

(A) 111.9 ohm

(B) 11.19 ohm

(C) 108 ohm

(D) 115 ohm
\end{tcolorbox}
\vspace{-3.5cm}
\end{table*}

\begin{table*}[h]
\caption{Optimized question for the Electrical Engineering task with the strategy applied.}
\vspace{-0.3cm}
\begin{tcolorbox}[
    colback=gray!5!white, 
    colframe=gray!75!black, 
    width=\textwidth, 
    boxrule=1.5pt,
    arc=2mm,
    left=10pt, right=10pt, top=8pt, bottom=8pt
]
\small
\textbf{Given:}

\quad • SCR gate cathode characteristic: straight line of 130

\quad • Triggered source volume: 15 V

\quad • Allowable gate power dissipation: 0.5 W

\textbf{Question:} Compute the gate source resistance.

Options:

(A) 111.9 ohm

(B) 11.19 ohm

(C) 108 ohm

(D) 115 ohm
\end{tcolorbox}
\end{table*}

\clearpage


\begin{table*}[t!]
\vspace{-1.0cm}
\caption{Optimized prompt for the Marketing task generated by Helix through dual-helix co-evolution.}
\vspace{-0.3cm}
\begin{tcolorbox}[
    colback=blue!5!white, 
    colframe=blue!75!black, 
    width=\textwidth, 
    boxrule=1.5pt,
    arc=2mm,
    left=10pt, right=10pt, top=8pt, bottom=8pt
]
\small
Answer marketing questions by carefully analyzing the question stem to identify which specific marketing concept, theory, or framework is being tested. Pay special attention to any theorist names, years, or distinguishing notes provided in parentheses after the question stem or options, as these indicate the specific theory or model being referenced.

When evaluating options, use the supplementary notes to distinguish between similar-sounding terms. For example, if options include ``(focuses on stages)'' versus ``(focuses on hierarchy),'' recognize that these notes highlight the key conceptual difference between otherwise similar-looking frameworks.

For questions about marketing theories or models, recall the key characteristics of each framework: What is its primary focus? What are its main components or stages? Who developed it and when? How does it differ from similar frameworks?

Systematically eliminate options that do not match the specific characteristics described in the question stem or that contradict the supplementary distinguishing notes. Select the option that most precisely matches both the theoretical content being tested and any specific attributes mentioned in parenthetical notes.
\end{tcolorbox}
\vspace{-5.0cm}
\end{table*}

\begin{table*}[h]
\caption{Optimized question strategy for the Marketing task generated by Helix through dual-helix co-evolution.}
\vspace{-0.3cm}
\begin{tcolorbox}[
    colback=green!5!white, 
    colframe=green!75!black, 
    width=\textwidth, 
    boxrule=1.5pt,
    arc=2mm,
    left=10pt, right=10pt, top=8pt, bottom=8pt
]
\small
\textbf{Strategy Type:} Formatting

\textbf{Rules:}
\begin{enumerate}[leftmargin=*, itemsep=4pt]
    \item \textbf{Primary rule:} For questions testing specific marketing theories or frameworks, add the theorist name and year in parentheses after the question stem (e.g., ``This model is called: (Rogers, 1962)'').
    \item \textbf{Secondary rule:} When options contain similar-sounding terms or concepts, add a brief distinguishing note in parentheses after each option to highlight the key difference (e.g., ``(focuses on stages)'' vs ``(focuses on hierarchy)'').
    \item \textbf{Preservation rule:} Keep all original question text, numbers, and core option content unchanged. Only add the supplementary notes.
\end{enumerate}
\end{tcolorbox}
\vspace{-5.0cm}
\end{table*}

\begin{table*}[h]
\caption{Original question from the Marketing task before optimization.}
\vspace{-0.3cm}
\begin{tcolorbox}[
    colback=gray!10!white, 
    colframe=gray!60!black, 
    width=\textwidth, 
    boxrule=1.5pt,
    arc=2mm,
    left=10pt, right=10pt, top=8pt, bottom=8pt
]
\small
This is a hierarchy of effects or sequential model used to explain how advertising works:

Options:

(A) ADD

(B) AIDA

(C) PESTLE

(D) SWOT
\end{tcolorbox}
\vspace{-5.0cm}
\end{table*}

\begin{table*}[h]
\caption{Optimized question for the Marketing task with the strategy applied.}
\vspace{-0.3cm}
\begin{tcolorbox}[
    colback=gray!5!white, 
    colframe=gray!75!black, 
    width=\textwidth, 
    boxrule=1.5pt,
    arc=2mm,
    left=10pt, right=10pt, top=8pt, bottom=8pt
]
\small
This is a hierarchy of effects or sequential model used to explain how advertising works: (Developed by E. St. Elmo Lewis, 1898)

Options:

(A) ADD (focuses on attention deficit)

(B) AIDA (focuses on sequential stages: Attention → Interest → Desire → Action)

(C) PESTLE (environmental analysis framework)

(D) SWOT (strategic planning tool)
\end{tcolorbox}
\end{table*}

\clearpage


\begin{table*}[t!]

\caption{Optimized prompt for the History task generated by Helix through dual-helix co-evolution.}
\vspace{-0.3cm}
\begin{tcolorbox}[
    colback=blue!5!white, 
    colframe=blue!75!black, 
    width=\textwidth, 
    boxrule=1.5pt,
    arc=2mm,
    left=10pt, right=10pt, top=8pt, bottom=8pt
]
\small
You are an expert historian analyzing primary and secondary historical sources. When answering history questions:

\textbf{Step 1: Carefully Read the Source Material} - Identify the document type (speech, treaty, letter, excerpt, etc.). Note the author, date, and historical context. Highlight key phrases that reveal the author's perspective or purpose.

\textbf{Step 2: Analyze the Question} - Determine what the question is asking (inference, comparison, causation, impact, etc.). Identify keywords that connect to the source material. Consider what historical knowledge is required.

\textbf{Step 3: Evaluate Each Option Systematically} - Eliminate options that contradict the source. Eliminate options that are historically inaccurate. Look for evidence in the text that directly supports or refutes each option. Be wary of options that are partially correct but incomplete.

\textbf{Step 4: Apply Historical Reasoning} - Consider cause and effect relationships. Recognize historical patterns and precedents. Understand the broader historical context. Distinguish between correlation and causation.

\textbf{Critical Reminders:} The correct answer MUST be supported by evidence in the source text. Avoid presentism. When comparing historical events, look for structural similarities. Pay attention to the specific wording of the question.

Now, carefully analyze the question and select the most accurate answer based on the evidence and your historical expertise.
\end{tcolorbox}
\vspace{-0cm}
\end{table*}

\begin{table*}[h]
\caption{Optimized question strategy for the History task generated by Helix through dual-helix co-evolution.}
\vspace{-0.3cm}
\begin{tcolorbox}[
    colback=green!5!white, 
    colframe=green!75!black, 
    width=\textwidth, 
    boxrule=1.5pt,
    arc=2mm,
    left=10pt, right=10pt, top=8pt, bottom=8pt
]
\small
\textbf{Strategy Type:} Structuring

\textbf{Rules:}
\begin{enumerate}[leftmargin=*, itemsep=4pt]
    \item \textbf{Primary:} Extract and present document metadata (type, author, date, context) at the beginning, followed by key quoted evidence from the source, then the question with explicit type labeling.
    \item \textbf{Secondary:} Organize long source documents into ``\textbf{Document Overview}'' (metadata) and ``\textbf{Key Evidence}'' (2-3 critical quotes) sections to reduce information overload.
    \item \textbf{Preservation:} Maintain all original source text, questions, and options exactly as given; only reorganize structure without changing any wording.
\end{enumerate}
\end{tcolorbox}
\vspace{-0cm}
\end{table*}

\begin{table*}[h]
\caption{Original question from the History task before optimization.}
\vspace{-0.3cm}
\begin{tcolorbox}[
    colback=gray!10!white, 
    colframe=gray!60!black, 
    width=\textwidth, 
    boxrule=1.5pt,
    arc=2mm,
    left=10pt, right=10pt, top=8pt, bottom=8pt
]
\small
While the Inca are often noted for their territorial expansion, the causes of their many wars included more than conquest and resource acquisition. Military actions also served to unify diverse populations by creating a common enemy, build loyalty among subjects, and showcase military strength to both allies and rivals. The Inca leadership sometimes used warfare as an instrument to identify and promote capable commanders, strengthening internal administration and military organization.

One of the less obvious reasons for the Inca to fight so many wars may have been:

Options:
...
\end{tcolorbox}
\vspace{-0cm}
\end{table*}

\begin{table*}[h]
\caption{Optimized question for the History task with the strategy applied.}
\vspace{-0.3cm}
\begin{tcolorbox}[
    colback=gray!5!white, 
    colframe=gray!75!black, 
    width=\textwidth, 
    boxrule=1.5pt,
    arc=2mm,
    left=10pt, right=10pt, top=8pt, bottom=8pt
]
\small
\textbf{=== SOURCE DOCUMENT ===}

\textbf{Document Overview:}

\quad • Type: Historical summary

\quad • Author: Modern historian specializing in Inca Empire

\quad • Historical Context: The Inca Empire (c. 1438–1533 CE)

\textbf{Key Evidence:}

``While the Inca are often noted for their territorial expansion, the causes of their many wars included more than conquest and resource acquisition. Military actions also served to unify diverse populations by creating a common enemy, build loyalty among subjects, and showcase military strength to both allies and rivals.''

``The Inca leadership sometimes used warfare as an instrument to identify and promote capable commanders, strengthening internal administration and military organization.''

\textbf{=== QUESTION ===}

\textbf{Question Type:} Inference about less obvious motivations for war

One of the less obvious reasons for the Inca to fight so many wars may have been:

Options:
...
\end{tcolorbox}
\end{table*}

\clearpage


\begin{table*}[t!]

\caption{Optimized prompt for the Philosophy task generated by Helix through dual-helix co-evolution.}
\vspace{-0.3cm}
\begin{tcolorbox}[
    colback=blue!5!white, 
    colframe=blue!75!black, 
    width=\textwidth, 
    boxrule=1.5pt,
    arc=2mm,
    left=10pt, right=10pt, top=8pt, bottom=8pt
]
\small
You are answering a philosophy question. Follow these steps systematically to maximize accuracy.

\textbf{Step 1:} Identify Core Concept - Extract the precise philosophical concept, logical structure, or theoretical claim being tested in the question.

\textbf{Step 2:} Recognize Key Terminology - Identify diagnostic terms that signal specific doctrines, thinkers, or logical patterns.

\textbf{Step 3:} Activate Domain Knowledge - Recall relevant philosophical principles, logical rules, or theoretical positions associated with the identified concept.

\textbf{Step 4:} Analyze Each Option - Test each option against the core concept for logical consistency, theoretical accuracy, and conceptual precision.

\textbf{Step 5:} Identify Common Traps - Watch for conflated positions, misapplied logical rules, or intuitively appealing but technically incorrect answers.

\textbf{Step 6:} Apply Hierarchical Elimination - Rule out options with clear errors first, then apply strict verification to remaining candidates.

\textbf{Step 7:} Verify Final Selection - Confirm your choice precisely matches the concept's definition, follows valid principles, and represents authentic positions.

Provide your complete reasoning and final answer.
\end{tcolorbox}
\vspace{-0.5cm}
\end{table*}

\begin{table*}[h]
\caption{Optimized question strategy for the Philosophy task generated by Helix through dual-helix co-evolution.}
\vspace{-0.3cm}
\begin{tcolorbox}[
    colback=green!5!white, 
    colframe=green!75!black, 
    width=\textwidth, 
    boxrule=1.5pt,
    arc=2mm,
    left=10pt, right=10pt, top=8pt, bottom=8pt
]
\small
\textbf{Strategy Type:} Structuring

\textbf{Rules:}
\begin{enumerate}[leftmargin=*, itemsep=4pt]
    \item \textbf{Primary rule:} For arguments containing logical reasoning, format multi-step reasoning as ``P1 $\rightarrow$ P2 $\rightarrow$ C'' or display conditional structures as ``IF P1 AND P2 THEN C''.
    \item \textbf{Secondary rule:} Extract conditional relationships and display as: ``Conditional: IF X THEN Y'' for implications, ``Necessary: X requires Y'' for necessary conditions, ``Sufficient: X guarantees Y'' for sufficient conditions, ``IF AND ONLY IF'' for biconditionals.
    \item \textbf{Preservation rule:} Keep all original philosophical terminology, thinker names, symbolic notation, and option content exactly as given. Do not add solving hints or alter the question's meaning.
\end{enumerate}
\end{tcolorbox}
\vspace{-0.5cm}
\end{table*}

\begin{table*}[h]
\caption{Original question from the Philosophy task before optimization.}
\vspace{-0.3cm}
\begin{tcolorbox}[
    colback=gray!10!white, 
    colframe=gray!60!black, 
    width=\textwidth, 
    boxrule=1.5pt,
    arc=2mm,
    left=10pt, right=10pt, top=8pt, bottom=8pt
]
\small
Statement: All birds live in some nest.

Predicate notation: Bx = x is a bird; Ny = y is a nest; Lxy = x lives in y

Select the best translation into predicate logic:

Options:
...
\end{tcolorbox}
\vspace{-0.5cm}
\end{table*}

\begin{table*}[h]
\caption{Optimized question for the Philosophy task with the strategy applied.}
\vspace{-0.3cm}
\begin{tcolorbox}[
    colback=gray!5!white, 
    colframe=gray!75!black, 
    width=\textwidth, 
    boxrule=1.5pt,
    arc=2mm,
    left=10pt, right=10pt, top=8pt, bottom=8pt
]
\small
\textbf{Statement:} All birds live in some nest.

\textbf{Predicate notation:} Bx = x is a bird; Ny = y is a nest; Lxy = x lives in y

\textbf{Logical dependency:}

\textbf{Conditional:} IF Bx (x is a bird) THEN ($\exists$y)(Ny $\cdot$ Lxy) (there exists a nest y such that x lives in y)

Select the best translation into predicate logic:

Options:
...
\end{tcolorbox}
\end{table*}

\clearpage


\begin{table*}[t!]

\caption{Optimized prompt for the AQuA-RAT task generated by Helix through dual-helix co-evolution.}
\vspace{-0.3cm}
\begin{tcolorbox}[
    colback=blue!5!white, 
    colframe=blue!75!black, 
    width=\textwidth, 
    boxrule=1.5pt,
    arc=2mm,
    left=10pt, right=10pt, top=8pt, bottom=8pt
]
\small
You are solving a math multiple-choice problem. Follow these steps systematically to maximize accuracy.

\textbf{Step 1:} Extract Key Information - Identify all given numbers, variables, relationships, and constraints from the problem. List them explicitly.

\textbf{Step 2:} Determine Problem Type - Classify the problem: arithmetic, algebra, geometry, probability, number theory, or word problem with rates/ratios.

\textbf{Step 3:} Set Up the Solution Framework - Based on the problem type, establish the appropriate equation, formula, or logical approach before calculating.

\textbf{Step 4:} Execute Calculations Carefully - Perform step-by-step calculations. Show intermediate results. Double-check arithmetic operations.

\textbf{Step 5:} Verify Against Constraints - Confirm your answer satisfies all conditions stated in the problem. Check units and reasonableness.

\textbf{Step 6:} Match with Options - Compare your calculated result with the given options (A-E). If no exact match, identify the closest or re-examine your work.

\textbf{Step 7:} Eliminate and Confirm - Rule out options that violate constraints or produce impossible values. Verify your final choice is mathematically sound.

Provide your complete reasoning, then state your final answer as: Answer: (X)
\end{tcolorbox}
\vspace{-0.4cm}
\end{table*}

\begin{table*}[h]
\caption{Optimized question strategy for the AQuA-RAT task generated by Helix through dual-helix co-evolution.}
\vspace{-0.3cm}
\begin{tcolorbox}[
    colback=green!5!white, 
    colframe=green!75!black, 
    width=\textwidth, 
    boxrule=1.5pt,
    arc=2mm,
    left=10pt, right=10pt, top=8pt, bottom=8pt
]
\small
\textbf{Strategy Type:} Structuring

\textbf{Rules:}
\begin{enumerate}[leftmargin=*, itemsep=4pt]
    \item \textbf{Primary rule:} Extract all numerical values from the question and assign them meaningful variable names at the start. Format: ``Let: v = 60 km/hr, t = 9 sec, n = 5 items''. Use intuitive variable names (v for velocity, t for time, r for rate, p for price, n for count, etc.). Place this ``Let:'' line before the question text.
    \item \textbf{Secondary rule:} Add a ``Find:'' line to explicitly state what needs to be calculated. Format: ``Find: L (length of train)'' or ``Find: T (total time)''. Use a descriptive variable name with explanation in parentheses.
    \item \textbf{Preservation rule:} Keep the original question text completely intact after the variable extraction. Do not modify any numbers, wording, or option values in the original content.
\end{enumerate}
\end{tcolorbox}
\vspace{-0.4cm}
\end{table*}

\begin{table*}[h]
\caption{Original question from the AQuA-RAT task before optimization.}
\vspace{-0.3cm}
\begin{tcolorbox}[
    colback=gray!10!white, 
    colframe=gray!60!black, 
    width=\textwidth, 
    boxrule=1.5pt,
    arc=2mm,
    left=10pt, right=10pt, top=8pt, bottom=8pt
]
\small
A car is being driven, in a straight line and at a uniform speed, towards the base of a vertical tower. The top of the tower is observed from the car and, in the process, it takes 10 minutes for the angle of elevation to change from 45° to 60°. After how much more time will this car reach the base of the tower?

Options:

(A) 5($\sqrt{3}$ + 1)

(B) 6($\sqrt{3}$ + $\sqrt{2}$)

(C) 7($\sqrt{3}$ – 1)

(D) 8($\sqrt{3}$ – 2)

(E) None of these
\end{tcolorbox}
\vspace{-0.4cm}
\end{table*}

\begin{table*}[h]
\caption{Optimized question for the AQuA-RAT task with the strategy applied.}
\vspace{-0.3cm}
\begin{tcolorbox}[
    colback=gray!5!white, 
    colframe=gray!75!black, 
    width=\textwidth, 
    boxrule=1.5pt,
    arc=2mm,
    left=10pt, right=10pt, top=8pt, bottom=8pt
]
\small
\textbf{Let:} $\theta_1$ = 45° (initial angle of elevation), $\theta_2$ = 60° (final angle of elevation), $t_{12}$ = 10 minutes (time taken for angle change)

\textbf{Find:} $t_b$ (time remaining for the car to reach the base of the tower)

A car is being driven, in a straight line and at a uniform speed, towards the base of a vertical tower. The top of the tower is observed from the car and, in the process, it takes 10 minutes for the angle of elevation to change from 45° to 60°. After how much more time will this car reach the base of the tower?

Options:

(A) 5($\sqrt{3}$ + 1)

(B) 6($\sqrt{3}$ + $\sqrt{2}$)

(C) 7($\sqrt{3}$ – 1)

(D) 8($\sqrt{3}$ – 2)

(E) None of these
\end{tcolorbox}
\label{tab:aqua_optimized}
\end{table*}

\clearpage


\end{document}